\documentclass{nature}
\usepackage{cite}
\usepackage{adjustbox}
\usepackage{xcolor}
\usepackage{graphicx}
\usepackage{hyperref}
\usepackage{amssymb}
\usepackage{amsmath}
\usepackage{lineno}
\usepackage[normalem]{ulem}
\usepackage{tablefootnote}
\usepackage{longtable}
\makeatletter
\let\saved@includegraphics\includegraphics
\AtBeginDocument{\let\includegraphics\saved@includegraphics}
\renewenvironment*{figure}{\@float{figure}}{\end@float}
\makeatother
\newcommand\aap{Astron. Astrophys.}                
\newcommand\apj{Astrophys. J.}                 
\newcommand\apjl{Astrophys. J. Lett.}                
\newcommand\mnras{Mon. Not. R. Astron. Soc.}             

\title{Optical and ultraviolet pulsed emission from an accreting millisecond pulsar}

\author{F. Ambrosino$^{1,2,3,*,\dagger}$, A. Miraval Zanon$^{4,5,*, \dagger}$, A. Papitto$^{1}$, F. Coti Zelati$^{6,7,5}$, S. Campana$^{5}$, P.\,D'Avanzo$^{5}$, L. Stella$^{1}$, T. Di Salvo$^{8}$, L. Burderi$^{9}$, P. Casella$^{1}$, A. Sanna$^{9}$, D. de Martino$^{10}$, M.\,Cadelano$^{11,12}$, A.\,Ghedina$^{13}$, F. Leone$^{14}$, F. Meddi$^{3}$, P. Cretaro$^{15}$, M. C. Baglio$^{16,5}$, E. Poretti$^{13,5}$, R. P. Mignani$^{17,18}$, D. F. Torres$^{6,7,19}$, G. L. Israel$^{1}$, M. Cecconi$^{13}$, D. M. Russell$^{16}$, M. D. Gonzalez Gomez$^{13}$, A. L. Riverol Rodriguez$^{13}$, H. Perez Ventura$^{13}$, M. Hernandez Diaz$^{13}$, J. J. San Juan$^{13}$, D. M. Bramich$^{16}$, F. Lewis$^{20,21}$ 
}

\begin{document}

\maketitle

\begin{affiliations}
 \item INAF-Osservatorio Astronomico di Roma, Via Frascati 33, I-00078 Monte Porzio Catone (Roma), Italy
 \item INAF-Istituto di Astrofisica e Planetologia Spaziali, Via Fosso del Cavaliere 100, I-00133 Roma, Italy
 \item Sapienza Universit\`a di Roma, Piazzale Aldo Moro 5, I-00185 Roma, Italy
 \item Universit\`a dell'Insubria, Dipartimento di Scienza e Alta Tecnologia, Via Valleggio 11, I-22100 Como, Italy
 \item INAF-Osservatorio Astronomico di Brera, Via  Bianchi 46, I-23807 Merate (LC), Italy  
 \item Institute of Space Sciences (ICE, CSIC), Campus UAB, Carrer de Can Magrans, E-08193, Barcelona, Spain
 \item Institut d'Estudis Espacials de Catalunya (IEEC), Gran Capit\`a 2-4, E-08034, Barcelona, Spain
 \item Universit\`a degli Studi di Palermo, Dipartimento di Fisica e Chimica, via Archirafi 36, I-90123 Palermo, Italy
 \item Universit\`a degli Studi di Cagliari, Dipartimento di Fisica, SP Monserrato-Sestu km 0.7, I-09042 Monserrato, Italy
 
 \item INAF-Osservatorio Astronomico di Capodimonte, Salita Moiariello 16, 
 I-80131 Napoli, Italy
 
 \item Universit\`a di Bologna, Dipartimento di Fisica e Astronomia, Via Gobetti 93/2, I-40129 Bologna, Italy
 \item INAF-Osservatorio di Astrofisica e Scienze dello Spazio di Bologna, Via Gobetti 93/3, I-40129 Bologna, Italy
 \item Fundaci\'on Galileo Galilei - INAF, Rambla Jos\'e Ana Fern\'andez P\'erez 7, E-38712 Bre\~na Baja (TF), Spain 
 \item Universit\`a di Catania, Dipartimento di Fisica e Astronomia, Sezione Astrofisica, Via Santa Sofia 78, I-95123 Catania, Italy
 \item INFN-Sezione di Roma 1, Piazzale Aldo Moro 5, I-00185 Roma, Italy
 \item Center for Astro, Particle, and Planetary Physics, New York University Abu Dhabi, PO Box 129188, Abu Dhabi, UAE
\item INAF-Istituto di Astrofisica Spaziale e Fisica Cosmica Milano, Via A. Corti 12, I-20133, Milano, Italy
 \item Janusz Gil Institute of Astronomy, University of Zielona G\'ora, ul Szafrana 2, 65-265, Zielona G\'ora, Poland
 
 \item Instituci\'o Catalana de Recerca i Estudis Avan\c cats (ICREA), E-08010 Barcelona, Spain 
 \item Faulkes Telescope Project, School of Physics and Astronomy, Cardiff University, The Parade, Cardiff, CF24 3AA, Wales, UK
\item Astrophysics Research Institute, Liverpool John Moores University, 146 Brownlow Hill, Liverpool L3 5RF, UK\\ 
$*$These authors contributed equally to this work\\ 
$\dagger$ corresponding authors (e-mail: filippo.ambrosino@inaf.it, arianna.miraval@inaf.it)
\end{affiliations}


\begin{abstract}
Millisecond spinning, low magnetic field neutron stars are believed to attain their
fast rotation in a 0.1-1 Gyr-long phase during which they accrete matter 
endowed with angular momentum from a low-mass companion star\cite{Alpar}. 
Despite extensive searches, coherent periodicities originating from accreting neutron star 
magnetospheres have been detected only at X-ray energies\cite{Wijnands} and 
in $\sim$ 10\% of the presently known systems\cite{Campana2018}. 
Here we report the detection of optical and ultraviolet coherent pulsations at the X-ray period of the transient low mass X-ray binary system SAX\,J1808.4$-$3658, during an accretion outburst that occurred in August 2019\cite{Bult}.
At the time of the observations, the pulsar was surrounded by an accretion disc,  
displayed X-ray pulsations and its luminosity was consistent with 
magnetically funneled accretion onto the neutron star. 
Current accretion models fail to account for the luminosity of
both optical and ultraviolet pulsations; these are instead more likely driven by synchro-curvature radiation \cite{Torres, Harding2018} in the pulsar magnetosphere or just outside of it. This interpretation would imply that particle acceleration can take place even when mass accretion is going on, and opens up new perspectives in the study of coherent optical/UV pulsations from fast spinning accreting neutron stars in low-mass X-ray binary systems. 

\end{abstract}
Low mass X-ray binary (LMXB) systems hosting a weakly magnetic ($\sim 10^8$\,G) 
neutron star are believed to be 
progenitors of millisecond radio pulsars. The evolutionary link between 
the two classes was first demonstrated through the detection of fast coherent X-ray 
pulsations generated by accretion onto the neutron star magnetic poles and the ensuing 
lighthouse effect in several transient LMXBs\cite{Wijnands, Chakrabarty}. Definitive proof came with the 
discovery of a small group of {\it transitional} millisecond binary pulsars which  
alternate between rotationally-powered and accretion-powered states\cite{Archibald, Papitto2013}.
Fast coherent pulsations and their frequency evolution in accreting neutron star systems are a tool of fundamental importance to determine binary parameters and accretion torques, investigate the 
properties of disc-magnetosphere interaction and magnetically funneled accretion, and  
derive constraints on the equation of state of ultradense matter\cite{Watts}. 
Through the detection and precise determination of the spin and orbital ephemeris of LMXBs, especially the most luminous and closest ones, it is also possible to carry out {\it tuned}, increased sensitivity searches for gravitational waves at (twice) the neutron star rotational frequency.
Fast accretion-powered coherent pulsations have proven elusive: 
in more than three decades they were detected at X-ray energies in 22 \cite{Wijnands2006, Patruno, Campana2018} out of $\sim$190 LMXBs harbouring neutron stars\cite{Liu}, 
all of which are transient systems attaining peak luminosities of up to a 
several percents the 
Eddington limit. 
So far optical pulsations have been detected only from the transitional millisecond pulsar PSR\,J1023+0038\cite{Ambrosino}, during an X-ray sub-luminous disc state\cite{Papitto2019}. 
Both the X-ray and optical pulsations of this system, which happen almost exactly at unison, are believed to  originate from synchrotron radiation in the intrabinary shock just beyond the light cylinder radius, where the wind of relativistic particles ejected by the pulsar meets the accretion disc\cite{Papitto2019, Veledina, Campana2019}.


The transient low mass X-ray binary SAX\,J1808.4$-$3658 is the first-discovered
accreting millisecond X-ray pulsar (AMXP)\cite{Wijnands}. 
The pulsar spins with a period of 2.49\,ms and orbits a $\sim~0.04\ M_{\odot}$ companion star\cite{Wang} with a 2\,hr orbital period\cite{Chakrabarty}; it is located at a distance\cite{Galloway} of about 3.5\,kpc. 
Since its discovery in 1996, the source underwent nine $\sim 1$~month-long outbursts 
during which the X-ray source
luminosity\cite{Gilfanov} reached typically a few  10$^{36}$\,erg\,s$^{-1}$, starting from 
a quiescence level\cite{Stella2k} of $\sim 5\times10^{31}$\,erg\,s$^{-1}$.
The higher mass inflow rate giving rise to the X-ray outbursts causes also 
an increase in the source ultraviolet (UV) and
optical brightness by $\sim$ 4 and $\sim$ 3.5 magnitudes, as a result of enhanced 
irradiation of the companion star and outer disc regions 
\cite{Giles} by the X-rays from the inner disc region and the neutron star
\cite{Gilfanov, Gierlinski}.

\begin{figure}

\includegraphics[scale=0.35]{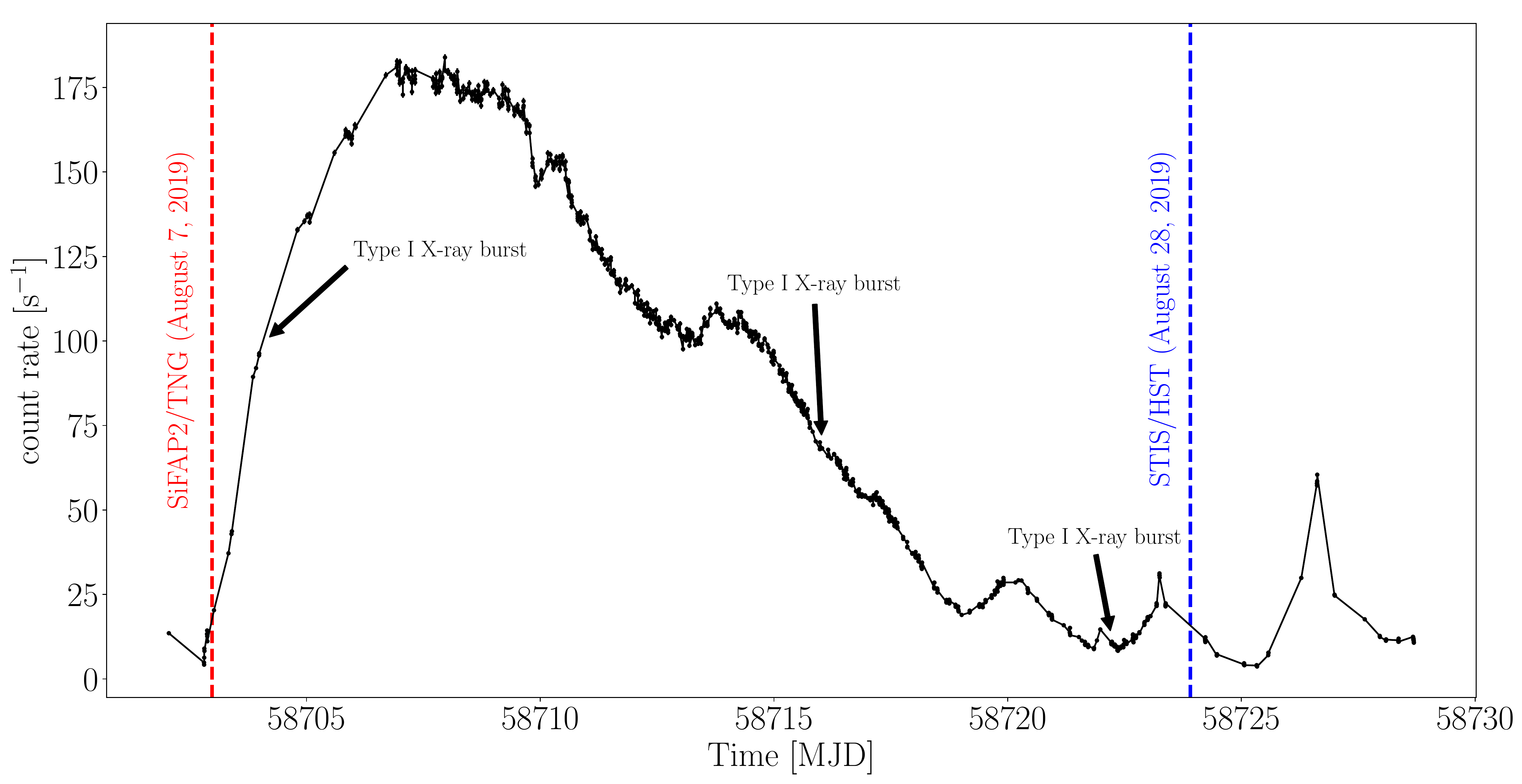}
\caption{\footnotesize XTI/NICER X-ray light curve (0.5--10\,keV) of the August 2019 outburst of  SAX\,J1808.4$-$3658. The red and the blue dashed lines indicate the epoch of our optical (August 7, 2019) and UV (August 28, 2019) fast-photometry observations, respectively. Intervals including type-I X-ray bursts have been removed from the plot. The epoch of their occurrence is indicated by arrows.}

\label{Fig1}
\end{figure}

In the summer of 2019, SAX\,J1808.4$-$3658 underwent another outburst\cite{Bult}, attaining a 
peak 0.5--10\,keV luminosity of $\sim 10^{36}$\,erg\,s$^{-1}$ on August 12, 
after a $\sim 5$~day rise. A decay followed and, starting from August 24, $\sim 4-5 $~day-long 
luminosity oscillations took place between $\sim 10^{34}$ and $\sim 10^{35}$\,erg\,s$^{-1}$.
Repeated observations with the X-ray Timing Instrument (XTI) on board the 
{\em Neutron Star Interior Composition Explorer} (NICER) closely 
monitored the evolution of the outburst and X-ray pulsations (see Fig.~\ref{Fig1}),  
yielding 
refined measurements of the neutron star spin period and orbital parameters (see Methods and \cite{Bult}). On August 7, during the rising phase of the outburst when the X-ray luminosity was 
$\sim 6\times 10^{34}$\,erg\,s$^{-1}$,  we observed the source for $\sim 1$~hr 
with the Silicon Fast Astronomical Photometer (SiFAP2)\cite{Ambrosino}, operating in the 320--900\,nm band
and mounted at the {\em Telescopio Nazionale Galileo} (TNG) in La Palma.
A $\sim$2~ks observation was carried out on August 28 with the Space Telescope Imaging Spectrograph (STIS), 
operating in $165-310$~nm UV band, 
on board the {\em Hubble Space Telescope} (HST), when the X-ray luminosity was $\sim 3.4\times 10^{34}$\,erg\,s$^{-1}$ in the final oscillating stages of the outburst (see Fig.~\ref{Fig1}).

\begin{figure}
\centering
\includegraphics[width=0.65\textwidth]{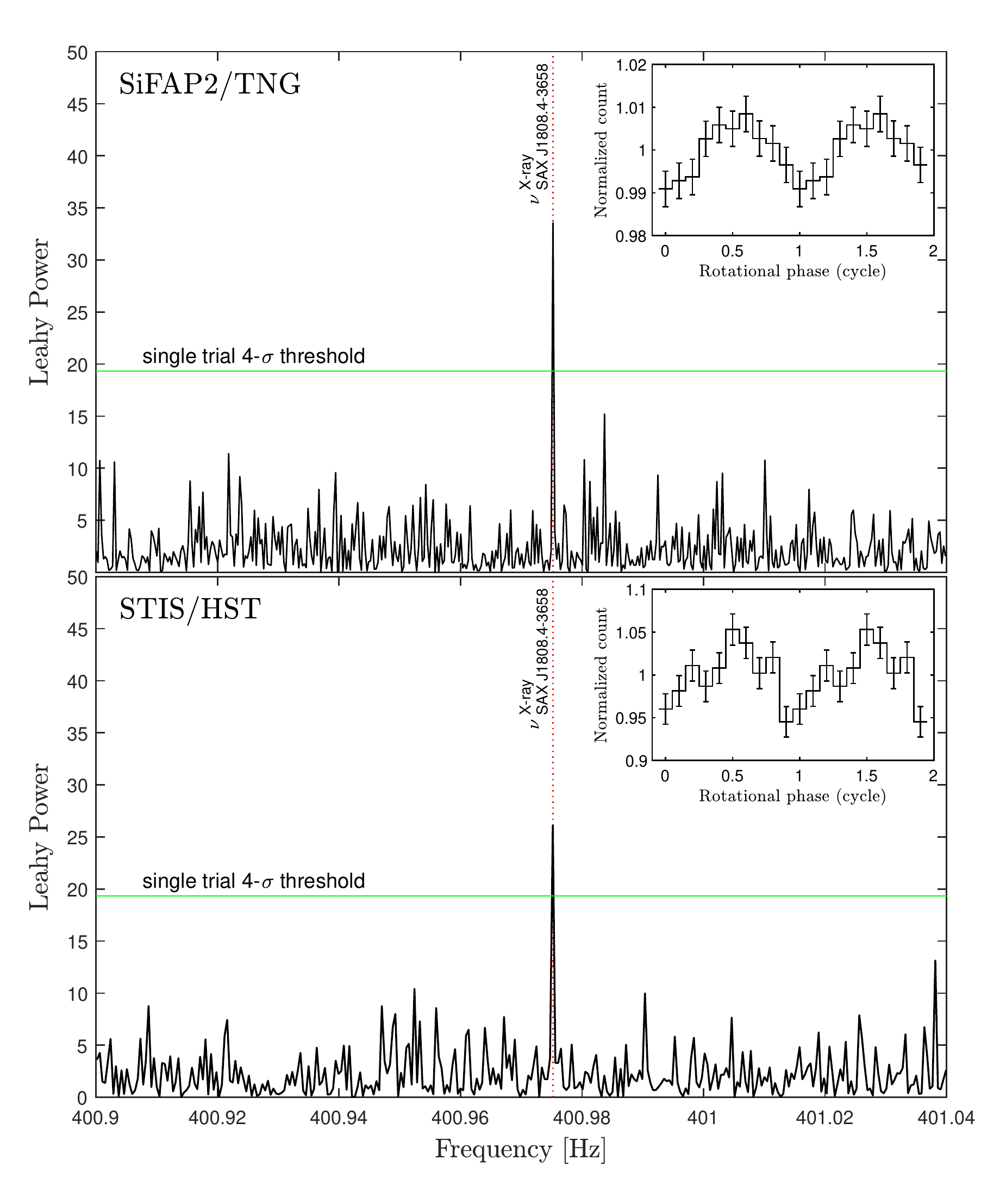}
\caption{\scriptsize 
Detection and shape of coherent optical and ultraviolet signals from SAX\,J1808.4$-$3658. Upper panel: Fourier power density spectrum 
of the optical (320--900\,nm) light curve from the 3.3~ks observation carried out with the SiFAP2 photometer mounted at the TNG, starting on August 7, 2019 at 22:31 (Coordinated Universal Time, UTC). Only a zoomed region around
the expected spin frequency of SAX\,J1808.4$-$3658 is shown.
Once corrected for the systematic drift of the SiFAP2 system clock (see Methods), photon arrival times were converted to the Solar System Barycentre (SSB) and then corrected for the pulsar orbital motion using the X-ray ephemeris reported in Table\ref{Tab1}. The light curve was binned at 100\,$\mu$s, corresponding to a Nyquist frequency of 5\,kHz. 
Lower panel: Fourier power density spectrum over the same frequency range
from the ultraviolet (165--310\,nm) light curve collected with STIS on board HST during a 2.2~ks observation starting on August 28, 2019 at 21:47 UTC. UV photon arrival times were first processed with the \textit{ODELAYTIME} task (see Methods) to shift them to the SSB, and then corrected for the pulsar orbital motion using the same X-ray ephemeris reported in Table\ref{Tab1}.  The light curve was rebinned to 500\,$\mu$s, yielding a Nyquist frequency of 1\,kHz. 
The dotted red vertical line marks the spin frequency of SAX\,J1808.4$-$3658 from the X-ray ephemeris, whose 
uncertainty on the spin frequency is small enough that only a single trial frequency has to be examined. The highest peaks in both panels coincide with this frequency, to within their Fourier resolution (see Methods).   
The green horizontal lines mark the power level corresponding to a probability of $6.3\times10^{-5}$ 
(4-$\sigma$) of being exceeded by white noise in a single frequency bin. 
The insets show the background-subtracted, normalised pulse profiles obtained by folding the optical and UV light curves at the X-ray period (Table\ref{Tab1}), two cycles are plotted for clarity.
Phases refer to the reference epoch of the SiFAP2 and STIS observations, respectively; errors bars are 1-$\sigma$.
}
\label{Fig2}
\end{figure}

The Fourier power density spectra of the high timing resolution optical and UV light curves are shown in
Fig.~\ref{Fig2}. In both cases, a narrow peak is present at the $\sim$ 401~Hz spin frequency 
of the neutron star, with a probability of random occurrence in a single frequency bin of  
$5.1\times10^{-8}$ and $2.3\times10^{-6}$ in the optical and UV data, respectively. 
The optical and UV pulsed light curves folded at the X-ray spin period reported in Table\ref{Tab1} and displaying a single-peaked quasi-sinusoidal profile are shown as insets in Fig.~\ref{Fig2}.
The background-subtracted rms amplitude of the optical pulsations was (0.55$\pm$0.06)\%, 
corresponding to  $L_{pulsed(opt)} \approx 2.7 \times 10^{31}$\,erg\,s$^{-1}$ (the total 325--690 nm optical 
luminosity was $L_{opt} \approx 5 \times 10^{33}$\,erg\,s$^{-1}$).  
In the XTI/NICER observations that covered the epoch of the
SiFAP2/TNG observation, the X-ray pulsations had a $\sim 9$ times larger rms amplitude (4.8$\pm$0.3\%), 
and a factor $\sim 100$ higher luminosity, $L_{pulsed(X)}\approx 2.3 \times 10^{33}$\,erg\,s$^{-1}$,
than the optical pulsations (see Methods for details).
Interestingly, the optical pulsation profile was shifted in phase by 
$\Delta\phi=0.55\pm0.02$ (or $\Delta\tau = 1.38\pm0.06$~ms in time) with respect to that in the X-rays, that is, virtually in anti-phase. We note that PSR\,J1023+0038 does not show such feature being its optical and X-ray pulse profiles almost in phase (time lag of $\sim 200\,\mu$s)\cite{Papitto2019}.

The UV coherent pulsations detected during the STIS/HST observation were 
(relatively) stronger than the optical pulsations from 3 weeks earlier: 
their (2.6$\pm$0.7)\% rms amplitude led to a pulsed 
165--310~nm luminosity of $L_{pulsed(UV)}=0.026~L_{UV} \approx 2 \times
10^{32}$\,erg\,s$^{-1}$. Correspondingly, the X-ray pulsations detected 
during the NICER observation carried out a few hours later had a (5.7$\pm$0.9)\% rms amplitude, 
and involved a factor of $\sim 10$ higher pulsed X-ray luminosity 
of $L_{pulsed(X)} \approx 1.9 \times 10^{33}$\,erg\,s$^{-1}$. 
Owing to the large uncertainties on the absolute timing of the HST data ($\sim1$\,s, HST helpdesk private communication) 
the relative phasing of the UV and X-ray profiles could not be determined. 

\begin{table*}

\caption{X-ray, UV and optical ephemeris of SAX\,J1808.4$-$3658 during the August 2019 outburst.}
\centering
\label{Tab1}      
\small                                      
\begin{adjustbox}{angle=90}                                      
\begin{tabular}{l l l l}          
\hline 

Parameter & X-ray & Ultraviolet  & Optical\\
\hline 
Right Ascension$^{a}$ ($\alpha$, J2000) & 18$^{\rm h}$ 08$^{\rm m}$ 27$^{\rm s}$.62 & -- & --\\
Declination$^{a}$ ($\delta$, J2000) & $-$36$^{\circ}$ 58$'$ 43.3$''$& -- & --\\
Validity Range [MJD] & $58702$--$58726$ & &  \\
Reference epoch $T_{ref}$ [MJD] & 58715.0 & -- & --\\
Time System & TDB & TDB & TDB \\
Planetary ephemeris & DE405 & DE200 & DE405 \\
Spin frequency ($\nu (T_{ref})$) [Hz]  & 400.975209660(9) & -- & --\\
Spin frequency ($\nu (T_{TNG})$)$^{b}$ [Hz] & 400.975210179(63) & --& 400.975225(72)$^\star$ \\ 
Spin frequency ($\nu (T_{HST})$)$^{c}$ [Hz] & 400.975209618(36) & 400.97517(10)$^\star$& -- \\
Spin frequency first derivative ($\dot{\nu}$) [Hz s$^{-1}$] & $ -(2.43\pm0.21) \times 10^{-13}$ & -- & --\\
Spin frequency second derivative ($\ddot{\nu}$) [Hz s$^{-2}$] & $ (4.9\pm1.1) \times 10^{-19}$ & -- & --\\
Orbital period ($P_b$) [s]&  7249.1572(14) & -- & --\\
Time of ascending node ($T^\star$) [MJD]&  58715.0220987(32) & -- & --\\ 
Projected semi-major axis [lt-s] &   0.0628099(35) & -- & --\\
$\chi^2$/d.o.f& 550/378 & -- & --\\
\hline 

\end{tabular}

\end{adjustbox}\\
$^{a}$ Values taken from \cite{Hartman2008}.\\
$^{b}$ Observation carried out on August 7, 2019 ($T_{\rm Start}^{\rm TNG}$ = 58702.9382176~MJD(UTC)) with SiFAP2/TNG.\\
$^{c}$ Observation carried out on August 28, 2019 ($T_{\rm Start}^{\rm HST}$ = 58723.9080081~MJD(UTC)) with STIS/HST. \\
$^\star$ Obtained with the epoch folding search technique.
\end{table*}
%


The presence of type I X-ray bursting activity as well as X-ray luminosities exceeding the spin-down power 
measured in quiescence\cite{Bult} ($1.6\times10^{34}$\,erg\,s$^{-1}$)
by up to two orders of magnitude testify that the outbursts of SAX\,J1808.4$-$3658 are powered by
mass accretion. 
Also at the time of the SiFAP2/TNG and STIS/HST observations, the X-ray luminosity was higher than the spin-down power by a factor 
of $\sim 2$ and 4 (though the pulsed X-ray luminosity was lower).
Moreover the source X-ray spectral and timing properties evolved moderately and continuously across the luminosity
swing of the outburst (as well as that of previous outbursts) down to $\sim 10^{34}$\,erg\,s$^{-1}$
without displaying any evidence for transitions to a non-accreting regime\cite{Bult} . 
The X-ray pulsations were likely generated by funneled accretion onto the magnetic pole.
Therefore, we conclude that the SAX\,J1808.4$-$3658 optical/UV pulsations detected nearly
simultaneously with the X-ray pulsations are the first detected certainly during the accretion phase of a millisecond spinning neutron star. In the following we discuss their possible origin.

Thermal emission from warm concentric rings surrounding the polar caps can be ruled out because, even considering a large emitting area ($\sim100$\,km$^2$), the temperature should attain unrealistically high values ($>$1\,MeV) to generate the observed optical and UV fluxes.
For the coherent signals not to be smeared out by light travel time 
delays, projection of their emission region along the line of sight should be smaller than $c P_{\rm spin}/2 \sim 300-400$\,km, where $c$ is the speed of light in vacuum. If the UV/optical pulsations arose from optically thick emission or 
reprocessing, the $\sim 2.7\times 10^{31}$\,erg\,s$^{-1}$ optical pulsed luminosity would imply a temperature of $\approx 2 (l/300\,{\rm km})^{-2}$\,keV, where $l$ is the size 
of the emitting region and a bolometric luminosity 
of $\approx 10^{41} (l/300\,{\rm km})^{-6}$\,erg\,s$^{-1}$, making this model untenable.
Moreover, the maximum size of this region limits the amount of energy that can be re-emitted for reprocessing of X-ray pulsations in the outer region of the disc ($R_{\rm out} \sim (2-3) \times 10^{10}$\,cm) to $\sim 10^{29}$\,erg\,s$^{-1}$ in the 320$-$900~nm band, even in the most favorable case of an inclination of 90$^{\circ}$.
The problem with this interpretation would be exacerbated if reprocessing or energy release in optically 
thick matter occurred in regions of size comparable to 
the characteristic scales of SAX\,J1808.4$-$3658, namely the neutron star radius
($R_{\rm NS} \sim 10$\,km), the inner disc boundary close to the corotation radius 
($r_c = (GM_{\rm NS}P_{\rm spin}^2/4\pi^2)^{1/3} \sim 32$\,km) or the light cylinder radius ($r_{lc} = cP_{\rm spin}/2\pi \sim 120$\,km); in fact all these regions 
are smaller than $\sim 300$\,km.



Hot electrons in the post-shock region of the accretion column  
will emit cyclotron photons at a fundamental energy of $E_{\rm cyc} \sim 4 (r/R_{\rm NS})^{-3}$\,eV,
for a surface magnetic field 
of SAX\,J1808.4$-$3658 of 
$B\sim 3.5 \times 10^8$\,G \cite{Burderi2006}. 
If the optically thick regime extended up to \emph{n}-th cyclotron harmonic, 
a Rayleigh-Jeans spectrum would result up to the corresponding 
energy\cite{Thompson}. 
For SAX\,J1808.4$-$3658 the maximum expected luminosity  would be
$L_{\rm cyc(opt)}\sim~10^{29}$\,erg\,s$^{-1}$ in the 320--900\,nm band,
and $L_{\rm cyc(UV)} \sim 6\times 10^{29}$\,erg\,s$^{-1}$ in the
165--310\,nm band (see Methods), i.e. more than two orders of magnitude lower 
than the measured values. Therefore 
it can also be excluded that self-absorbed cyclotron emission in the accretion column is 
responsible for the optical/UV pulsed flux of SAX\,J1808.4$-$3658, unless emission in these bands is strongly beamed, which is deemed unlikely given the high pulse duty cycle. Also, pencil beaming due to the reduction of the cyclotron opacity for photons propagating along the field lines\cite{Basko} is expected at energies much lower than $E_{\rm cyc}$, i.e. below the optical band within which we observed. These limitations would no longer hold if the optical/UV pulsed emission were produced by a coherent emission process\cite{Melrose2017}, whose specific intensity can vastly exceed that of thermal emission. However we note that coherent emission from rotation-powered pulsars is characterised by a steep power-law like radio spectrum and is not expected to operate at much higher frequencies.

Similar to the case of isolated rotation-powered pulsars\cite{Pacini}, synchro-curvature radiation\cite{Harding2018, Torres} by relativistic electrons and positrons accelerated by the 
rotating neutron star magnetosphere might give rise to the optical and UV pulsations of SAX\,J1808.4$-$3658. In this interpretation, the efficiency with which SAX\,J1808.4$-$3658 converts the spin-down power into pulsed UV and optical luminosity would be
$\eta_{UV}~\sim 1 \times 10^{-2}$ and $\eta_{opt}\sim 6 \times 10^{-4}$, respectively, the former being about 100 times larger than that of the Crab pulsar in the UV (165--310\,nm) and $B$ bands\cite{Mignani} (see Methods).
Such a high efficiency is much larger than that usually observed for isolated rotation-powered pulsars, and points toward the existence of a synergistic physical process.

Models based on magneto-hydrodynamic simulations\cite{Parfrey2017_2} predict that the neutron star magnetic field lines coupled to the disc within the corotation radius are rapidly twisted, pushed outwards and forced to open\cite{Parfrey2016}, a  phenomenon possible only in AMXPs with high magnetic diffusivity discs.
In this picture the rotation-powered mechanism would not be inhibited by the presence of the accretion disc; rather its power would increase (as compared to discless pulsars) owing to the opening of additional magnetic field lines and the corresponding flux enhancement across the light cylinder surface. This leads to an increase of the spin-down torque applied to the neutron star and a stronger electromagnetic pulsar wind\cite{Parfrey2017_2}. A net spin-down rate of order $\sim -1 \times 10^{-13}$\,Hz\,s$^{-1}$ would be expected in the case of SAX\,J1808.4--3658.
We note that a comparably large spin-down torque and rate may arise from magnetic field lines threading the disc beyond the corotation radius\cite{Kluziniak2007}.
If the high optical and UV pulsed luminosities of SAX\,J1808.4$-$3658  result from an enhanced 
rotation-powered mechanism, this must co-exist (or alternate on timescale shorter than those required to detect pulsations with current instrumentation) with the accretion-powered mechanism that produces the X-ray pulsations. 
Alternatively, the power of the so-called {\it striped wind}\cite{Coroniti} may also be enhanced by the 
same disc-magnetosphere interaction. In this model, part of the pulsar spin-down power is carried away in the form of low-frequency waves consisting of stripes of toroidal magnetic field; these structures propagate along the equatorial plane of the pulsar and are converted into a wind of relativistic magnetized plasma beyond the light cylinder radius through magnetic reconnection. In this framework, the optical and UV pulsations from SAX\,J1808.4$-$3658 would arise from beamed synchrotron radiation by heated charged particles moving 
close to the light cylinder radius\cite{Kirk2002}. Here,
synchrotron radiation is optically thin to emission in the UV and optical bands, 
as synchrotron self-absorption occurs below\cite{Rybicki} $E_{\rm break}\sim$ 0.04\,eV. 
Optical/UV pulsed emission from synchrotron radiation is still expected at distances  $\lesssim$ 600\,km from the neutron star, considering that the synchrotron cooling timescale is shorter than $P_{\rm spin}/2$. The efficiency of this process could be enhanced at the termination shock between the pulsar wind and the accretion disc\cite{Papitto2019, Veledina, Campana2019}.

An intriguing feature is the half cycle shift between the optical and the X-ray pulsations from SAX\,J1808.4$-$3658. 
It is tempting to consider the possibility that matter accretion takes place only on one pole of the neutron star, thus giving rise to the pulsed X-ray signal, whereas accretion is 
inhibited on the opposite side and a rotation-powered mechanism gives rise to the optical/UV pulsation in anti-phase. 
A dipole magnetic field whose center is shifted from the center of the neutron star might make this possible.  


The discovery of optical and UV pulsations during the accretion outburst of a millisecond X-ray pulsar shows that particle acceleration mechanisms can occur even in regimes where the magnetosphere is engulfed with accreting plasma for at least a significant fraction of the time. Moreover, it opens a novel observational window in the study of accreting neutron stars in low mass X-ray binaries, as 
the higher sensitivity afforded by optical and UV fast photometric observations may 
allow the discovery of coherent pulsations in sources and regimes for which X-ray pulsations 
have remained undetected. 


\begin{methods}

\subsection{TNG optical observation.}
The optical dataset of SAX\,J1808.4$-$3658 was collected with the Silicon Fast Astronomical Photometer (SiFAP2\cite{Ghedina}, TNG Director Discretionary Time, PI Papitto) mounted at the Nasmyth A focus of the INAF 3.58~m Telescopio Nazionale Galileo (TNG), located on the Roque de los Muchachos Observatory in La Palma (Canary Islands, Spain). SiFAP2, the upgraded version of SiFAP\cite{Meddi,Ambrosino2016}, is a two-channel ultra-fast photometer operating in the optical band (320--900\,nm) capable to tag the Time of Arrival (ToA) of individual photons with a time resolution of 8\,ns.
The absolute timing is provided by a commercial Global Positioning System (GPS) unit via the Pulse Per Second (PPS) signal with a nominal 25\,ns accuracy on the Universal Time Coordinated (UTC) worsened to less than 60\,$\mu$s because of the SiFAP2 electronics transfer function. This value was obtained from observations performed on the Crab pulsar for calibration purposes\cite{Papitto2019}. 
We carried out a single 3.3~ks observation of SAX\,J1808.4$-$3658 starting on August 7, 2019 at 22:31 (UTC), during the earliest stage of the outburst. The optical light curve of the source collected with SiFAP2 is shown in Fig.~\ref{LCs}. 
No filter was used during our run. The telescope elevation above the horizon was $\sim$ 24\,deg corresponding to an airmass of $\sim$ 2.5, while seeing conditions varied within the range from 0.5 up to 0.9\,arcsec (at the Zenith). The Moon was at an angular distance of 47\,deg from the target, increasing the background contribution by $\sim$ 70\%. During the acquisition, we also measured the sky background signal (taking into account also a dark count rate of 1.8 $\times$ 10$^3$\,s$^{-1}$ for the sensors) by moving the telescope 10\,arcsec away from the target towards the east direction twice during the observation, for about 30\,s each time. We obtained an average count rate of BKG$_{TNG}$ = 34953$\pm$86\,s$^{-1}$, representing a contribution of more than 90\% of the total count rate (R$_{TNG}$ = 38560.6$\pm$6.5~s$^{-1}$)  
collected by pointing the telescope in the direction of SAX\,J1808.4$-$3658. 
A reference star, TYC 7403$-$655$-$1 (RA = 18:07:56.38, DEC = $-$36:55:07.35, $V$ = 12.19~mag) located 432~arcsec away from SAX\,J1808.4$-$3658 was also simultaneously observed to monitor the atmospheric variations as well as to verify the absence of spurious periodic signals due to instrumental noise at the pulsar spin frequency. 
As it occurred also in previous observing runs, the SiFAP2 clock drifted by $\Delta t=5.2$\,ms with respect to the time measured by two Global Positioning System (GPS) pulse-per-second signals used to mark the beginning and the end of the observation. We corrected the arrival times assuming that the drift evolved following a linear function of time. This procedure already proved to be efficient in recovering the pulse frequency of both the Crab pulsar and the millisecond pulsar PSR J1023+0038\cite{Ambrosino, Papitto2019}. Laboratory tests showed that the thermal jitter of the SiFAP2 system clock could be safely neglected because its relative uncertainty for ms spin periods is several tens of times smaller than our measurements\cite{Ambrosino}.
The photon arrival times obtained in this way were then referred to the Solar System Barycenter (TDB time system) using the position of the optical counterpart provided by \cite{Hartman2008} and the geocentric location of the TNG (X = 5327447.4810\,m, Y = $-$1719594.9272\,m, Z = 3051174.6663\,m), along with the {\bf }JPL DE405 ephemeris. 

\subsection{Ultraviolet observation.}
We observed SAX\,J1808.4$-$3658 with the Space Telescope Imaging Spectrograph (STIS, GO/DD-15987, PI Miraval Zanon) on board the Hubble Space Telescope (HST) starting on 28 August 2019 at 21:47 (UTC) during the latest stage of the outburst. The UV light curve of the source acquired with STIS is shown in Fig.~\ref{LCs}.
The spectroscopic observation was performed in \textit{TIME-TAG} mode with 125\,$\mu$s time resolution for about 2.2\,ks by means of the NUV-MAMA detector. We used the G230L grating equipped with a 52$\times$0.2\,arcsec slit ensuring a spectral resolution of $\sim$500 over the nominal range (first order). 
The total count rate collected by the instrument was R$_{HST}$ = 2016.74$\pm$0.95~s$^{-1}$, with a background contribution of about 30\% (BKG$_{HST}$ = 653.36$\pm$0.54~s$^{-1}$). The background signal was estimated by selecting photons in the STIS slit channels outside the source region (see Timing analysis section) and averaging them. The resulting value was then normalised to the total number of slit channels.

\subsection{X-ray observations.} 
The X-ray Timing Instrument\cite{Gendreau2016} (XTI) on board the {\em Neutron Star Interior Composition Explorer}\cite{Gendreau2017} (NICER) observed the SAX\,J1808.4$-$3658 outburst\cite{Bult} from 30 July until 16 September 2019 for a total exposure time of 387.7\,ks. The events across the 0.2--12\,keV band
were processed and screened using HEASOFT version 6.28 
and NICERDAS version  7a. 
We applied standard cleaning and filtering criteria, selecting only the time intervals during which the pointing offset from the nominal source position was smaller than 0.015\,deg, the source was at least 30\,deg away from the Earth limb (at least 40\,deg in the case of a Sun-illuminated Earth) and the International Space Station was outside the South Atlantic Anomaly. The photon arrival times were corrected for the motion in the Solar System barycentre (TDB time system) using the position of the optical counterpart\cite{Hartman2008} and JPL DE405 ephemeris. The X-ray light curve of the source outburst from 2019, August 7 to August 31 is shown in Fig.~\ref{Fig1}. We removed Type-I X-ray bursts occurred in the time intervals 58704.81059-58704.81186\,MJD, 58716.08876-58716.09104\,MJD and 58722.41759-58722.41921\,MJD. 
\begin{figure}
\centering
\includegraphics[scale=0.75]{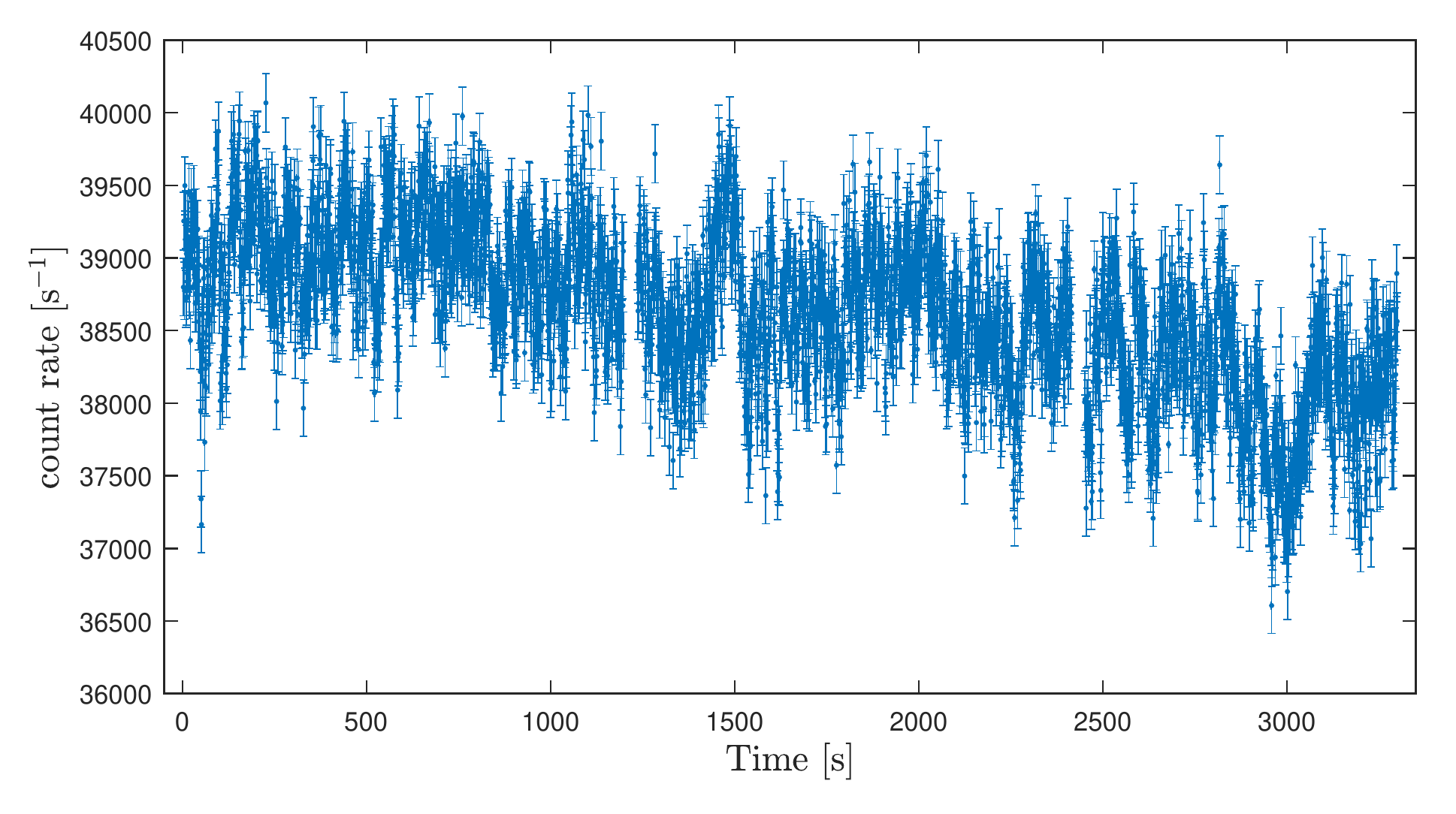}
\includegraphics[scale=0.75]{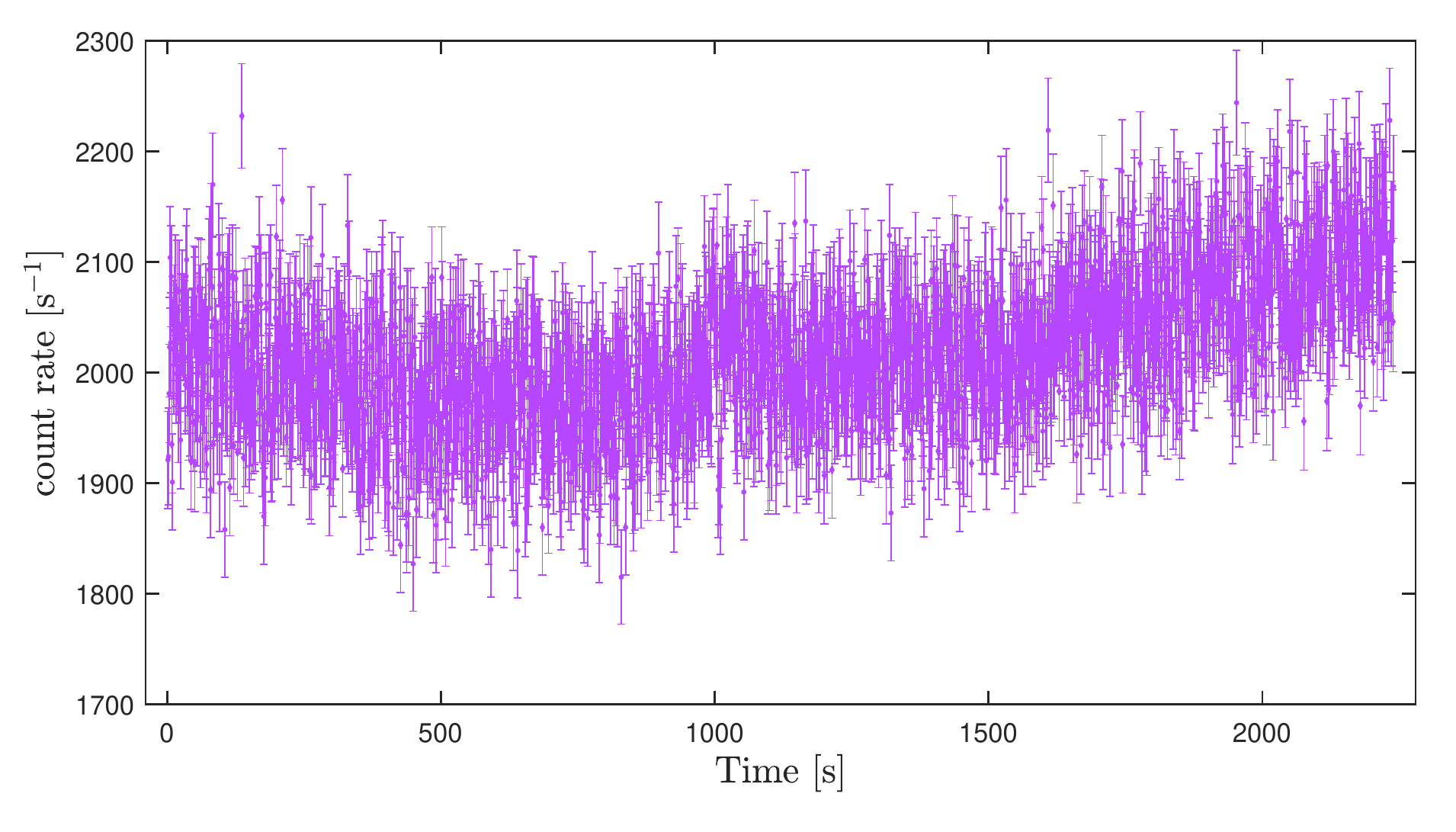}
\caption{\footnotesize Optical and UV light curves of SAX\,J1808.4$-$3658. \emph{Top}: Optical count rate of SAX\,J1808.4$-$3658 collected with SiFAP2 on August 7, 2019 during a total exposure time of 3.3\,ks. The time on the x-axis represents the elapsed time since 58702.9382176~MJD(UTC). 
The orbital phase interval is 0.04--0.49. \emph{Bottom}: Ultraviolet count rate of SAX\,J1808.4$-$3658 acquired with STIS on August 28, 2019 during a total exposure time of 2.2\,ks. The time on the x-axis represents the elapsed time since 58723.9080081~MJD(UTC). The observed orbital phase interval is -0.06--0.25. Both light curves are plotted with a bin time of 1\,s; error bars represent 1-$\sigma$ uncertainties.}
\label{LCs}
\end{figure} 
\subsection{Timing analysis.}

We measured the X-ray pulsar spin and orbital parameters by analysing the observations performed by NICER between August 7 and 31 (i.e., MJD 58702--58726). We corrected the arrival times using the orbital parameters previously measured\cite{Bult}. We folded 1~ks-long segments of NICER data in 16 phase bins around a preliminary estimate of the pulse period\cite{Bult}. We fitted the pulse profiles with a single sinusoidal component, modeled the evolution of the pulse phases with a function composed of a third-order polynomial and terms resulting from corrections to the orbital parameters (see, e.g.~\cite{Papitto2011}), and obtained the timing solution listed in Table~\ref{Tab1}. This solution is characterized by a $\chi^2$ of 550 for 378 degrees of freedom, indicating a formally unacceptable fit. However, even adopting higher-order polynomials, the fit quality did not improve significantly. No trend is apparent in the residuals shown in the bottom panel of Fig.~\ref{NICERtiming}, and we attribute the high fit reduced $\chi^2$ to the phase timing noise that is known to affect
the phases observed from this and other AMXPs\cite{Patruno}. The X-ray timing solution derived here is only aimed at performing a search for optical/UV pulses, and modelling such a timing noise component is beyond the scope of this paper. However, we note that Bult et al.\cite{Bult} derived a timing solution measuring the pulse phases in each continuous good time interval (generally longer than 1 ks) and considering either a second-order polynomial or a flux-adjusted phase model in an attempt to model the timing noise, and obtained similar values of the fit $\chi^2$ than that reported here. The different models used by those authors explain the slight differences between their ephemeris and the ones we obtained. In any case, we checked that our results do not change when using their ephemeris.

\begin{figure}
\centering
\includegraphics[scale=1.3]{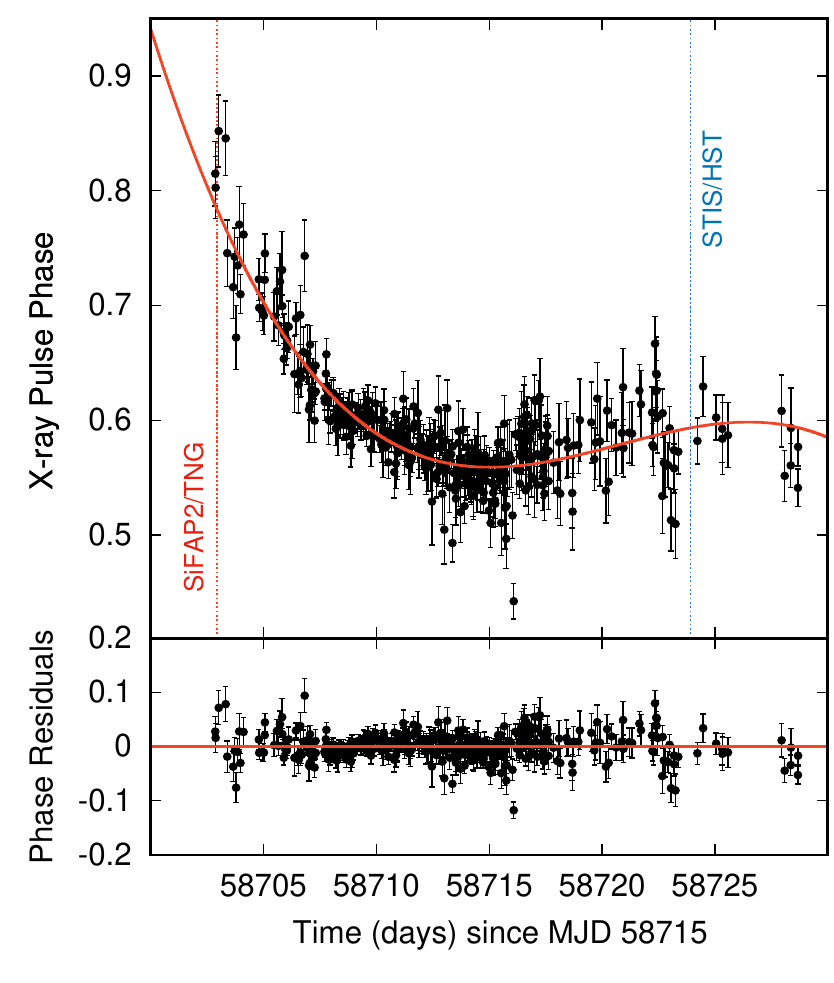}
\caption{\footnotesize Phase analysis of X-ray data from XTI/NICER. \emph{Top}: XTI/NICER X-ray pulse phases as a function of time. The red line shows the best fitting  third-order polynomial function. The dashed red and blue vertical lines mark the epochs of SiFAP2/TNG and STIS/HST observations. \emph{Bottom}: Residuals with respect to the best-fitting solution. Error bars represent 1-$\sigma$ uncertainties for both panels.
}
\label{NICERtiming}
\end{figure} 

Since the TNG observation lasted 3.3\,ks, the spacing of Fourier independent frequencies was $\delta\nu_{TNG}=3.0\times10^{-4}$\,Hz. This is $\sim$ 5000 times coarser than the uncertainties on the X-ray spin frequency evaluated at the epoch of the TNG observation ($\sigma_{\nu}(T_{TNG})=6.3\times10^{-8}$\,Hz, see Table\,\ref{Tab1}). After both the barycentric correction and the demodulation for the pulsar orbital motion had been applied, only a single trial frequency had to be searched in the TNG dataset to investigate the presence of a coherent signal at the same frequency as determined from the analysis of the X-ray data. We calculated the Fourier power density spectrum of the TNG light curve and measured a Leahy normalised\cite{Leahy83} power of $33.6$ at a frequency of $400.97522(15)$\,Hz. The single-trial probability associated to random white noise fluctuations is $p=5.1\times10^{-8}$. 
We note that no significant peak at the expected pulsar spin frequency was found out neither in the power density spectrum of the light curve in which orbital demodulation was not applied nor in the reference star light curve. A precise knowledge of the spin and orbital ephemeris were thus essential 
for detecting the optical (and UV) coherent modulation at the spin period. To refine the measurement of the frequency of the optical pulse, we performed an epoch folding search\cite{Leahy87} adopting 10 phase bins and a period stepsize of $\delta{P}_{TNG,EFS}=9.4\times10^{-11}$\,s
measuring a chi-squared value of $S_{max}=38.6$ with a corresponding best-fitting period of $P_{TNG,EFS}=0.00249391967(45)$\,s. The 1-$\sigma$ uncertainty reported in parentheses was computed\cite{Leahy87} by using the equation $\sigma_P=P^2/(2T_{exp})(S_{max}/(n-1)-1)^{-0.63}$, where $T_{exp}$ is the total exposure time of the TNG optical observation. In addition, we also performed the bin-free $Z_n^2$ test\cite{Buccheri} assuming a purely sinuosidal profile shape ($n = 1$), obtaining a value of $Z_1^2$ = 34.3 associated to a best-fitting folding period of $P_{TNG,Z_1^2}=0.00249391976(62)$\,s.
This value is compatible with that estimated from the X-ray data within the uncertainties. We reported the computed
chi-squared distribution with the best-fitting Gaussian model in Fig.~\ref{EFSs}.
We then folded the TNG optical light curve using the X-ray pulse parameters, and modeled the pulse profile obtained in this way (inset of Fig.~\ref{Fig2}) using a single sinusoidal component with a rms amplitude of $(0.051\pm0.005)\%$. Taking into account the background (see above), we then estimated the source rms amplitude as $A_{TNG}^{rms}=(0.55\pm0.06)\%$. Folding at the same spin frequency the SiFAP2/TNG light curve as well as the XTI/NICER light curve extracted over the time interval between August 7 at 19:18:49 and August 8 at 00:34:55 UTC (a subset of observation ids. 2050260109 and 2050260110) highlighted a phase difference of $\Delta\phi = (0.55\pm0.02)$, corresponding to a time lag of $\tau = (1.38\pm0.06$)\,ms (see Fig.~\ref{Fig_PulseProfiles}). \cite{Papitto2019} estimated a SiFAP2 absolute timing accuracy of $< 60\,\mu$s, whereas the corresponding NICER values is $< 0.3\,\mu$s, both much lower than the uncertainty affecting the lag measured. To estimate the effect of any residual relative timing uncertainty caused by, e.g., the uncertainty on the time dependence of the SiFAP2 clock drift, we assumed that the frequency of the optical and X-ray signals were exactly equal, and estimated the phase uncertainty driven by a frequency drift of an amount equal to the measurement error ($\sigma_\nu=5\times10^{-5}$\,Hz; see Table\,\ref{Tab1}), obtaining a maximum lag of 0.4\,ms. Future observations ensuring a full-orbit coverage will confirm the significance and magnitude of the pulse lag.

Hints for slight variations of the optical pulse amplitude were found, although their significance is low. The observed (i.e. not subtracted for the background) rms amplitude varied between (0.07$\pm$0.01)\% and (0.04$\pm$0.01)\% over 0.8~ks-long intervals. The TNG optical observation covered about one half the orbital period, from phase 0.04 to 0.49, that is, from shortly after ascending node (phase 0) to close to descending node (phase 0.5). 
The coherent signal was detected at all phases, although the maximum rms amplitude was detected when the 
pulsar was close to the ascending node.
Given the intrinsic weakness of the signal, the TNG optical observation was too short to allow us to
derive the orbital parameters of the optical pulse from a pulse timing analysis of that data set. To confirm the association of the optical coherent signal with the pulsar in SAX\,J1808.4$-$3658, we then determined the variation of the signal strength by varying the orbital parameters adopted in the correction of the photon arrival times with respect to the values measured from the analysis of the X-ray pulsations. We  rerun the epoch folding periodicity search of the light curves by allowing the epoch of the ascending node ($T^\star$), and
the projected semi-major axis ($x = a \sin{i}/c$) to vary over a grid of values spaced by $\delta T^\star = 2.5$~s and $\delta x = 1\times10^{-3}$~lt-s. The folding period was allowed to vary. 
The distributions of chi-squared values associated to the best folding period computed by varying independently $T^\star$ and $x$ are reported in Fig.~\ref{Fig_deltaTasc} and in Fig.~\ref{Fig_delta_asini}, respectively. We performed a Gaussian fitting on both the chi-squared distributions obtaining the position of the two centroids at $\Delta T^\star = -(4.4 \pm 2.3)$~s, and $\Delta (a sin(i)/c) = -(0.32 \pm 0.33)$~lt-ms. However, we caution that a coverage of an entire orbital cycle seems warranted to draw firm conclusions on the significance of a possible offset between the two pulse profiles.

We then analysed the ultraviolet events obtained from the observation performed with STIS. We corrected the position of slit channels thanks to an external custom function (\url{https://github.com/Alymantara/stis_photons}) and selected events (i.e. ToAs) belonging to channels of the slit within the 991$-$1005 (edges excluded) interval to isolate the source signal and minimise the background contribution. We also selected the 165--310\,nm wavelength interval to avoid noisy contribution due to the poor response of the G230L grating at the edge wavelengths. The list of good ToAs was corrected to the SSB by using the \textit{ODELAYTIME} task (subroutine available in the \textit{IRAF/STDAS} software package) and the JPL DE200 ephemeris.
We applied the same procedure as previously done for SiFAP2 data on the STIS dataset to search for the UV pulsed emission from SAX\,J1808.4$-$3658. After having demodulated the UV photons ToAs by correcting them for the R{\o}mer delays due to the orbital motion, we computed the Fourier power density spectrum. We found a Leahy normalised\cite{Leahy83} power of $26.1$ at a frequency of $400.97518(22)$\,Hz indicating coherent UV pulsations around the expected pulsar spin frequency with an associated single-trial probability of $2.3\times10^{-6}$. As in the case of the optical dataset, no significant peak at the expected pulsar spin frequency was found in the power density spectrum of the non-demodulated light curve. We then performed an epoch folding search using $n = 10$ phase bins and a period resolution of $\delta{P}_{HST,EFS}=1.4\times10^{-10}$\,s. 
We measured a chi-squared value of $S_{max}=39.6$, and a best-fitting period of $P_{HST,EFS}=0.00249391998(64)$\,s,
well in agreement, within the uncertainties, with the period from the X-ray data. The 1-$\sigma$ uncertainty reported in  parentheses was computed as in the optical dataset. We reported the computed chi-squared distribution with the best-fitting Gaussian model in Fig.~\ref{EFSs}. We performed the bin-free $Z_n^2$ test with a $n = 1$ component, deriving a value of $Z_1^2$ = 29.0 associated to a best-fitting folding period of $P_{HST,Z_1^2}=0.00249391997(98)$\,s.
We then folded the HST ultraviolet data at the best period obtained from the X-ray timing analysis (see Tab.~\ref{Tab1}), and plotted the background-subtracted pulse profile in the inset of Fig.~\ref{Fig2}. We described the shape of the UV modulation with a single Fourier component with rms fractional amplitude A$_{HST}^{rms}$ = (2.6$\pm$0.7)\%; note, however,
that the rise to the peaks of the modulation was somewhat slower than the decay.

\begin{figure}
\centering
\includegraphics[scale=0.7]{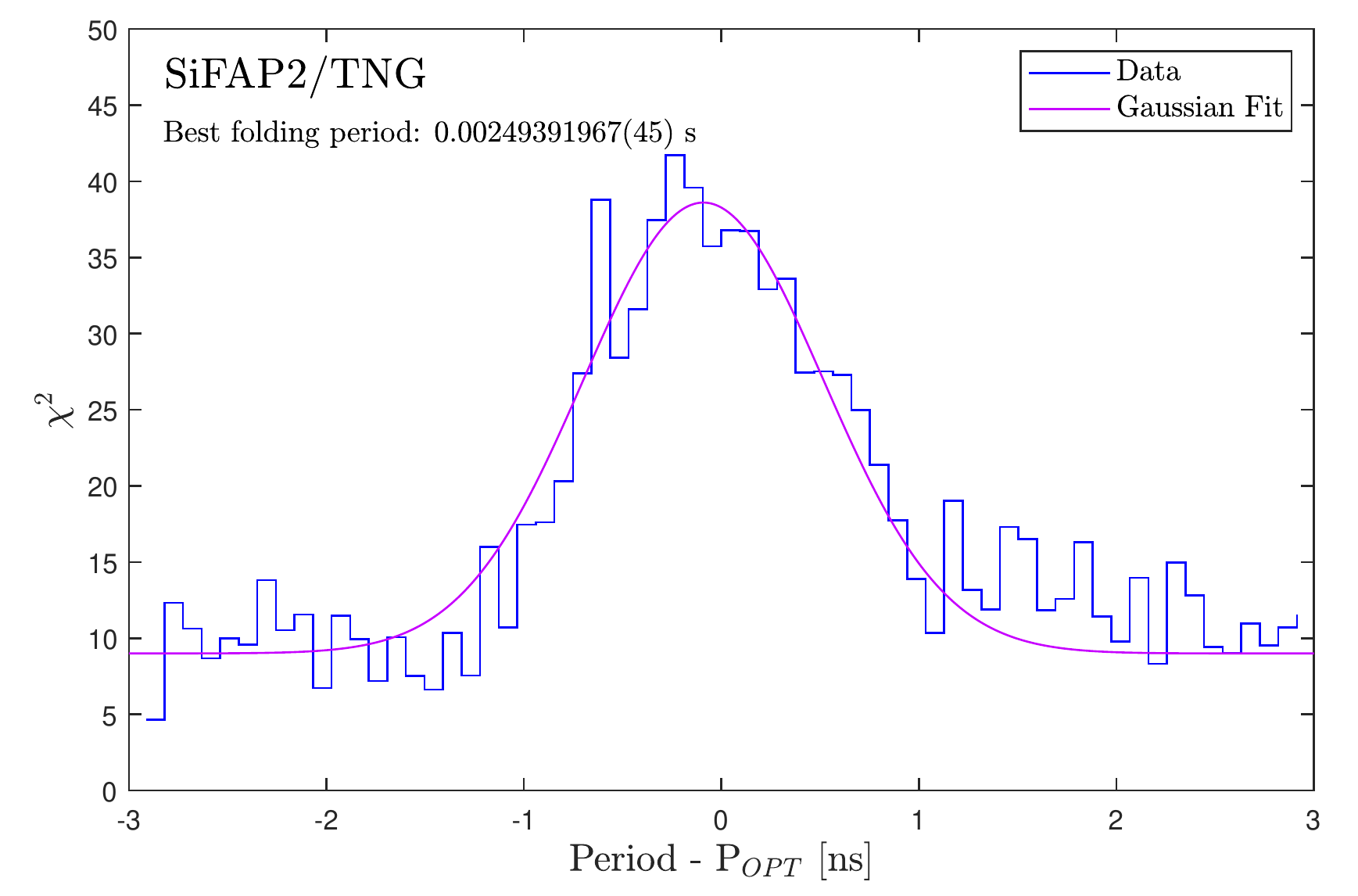}
\includegraphics[scale=0.7]{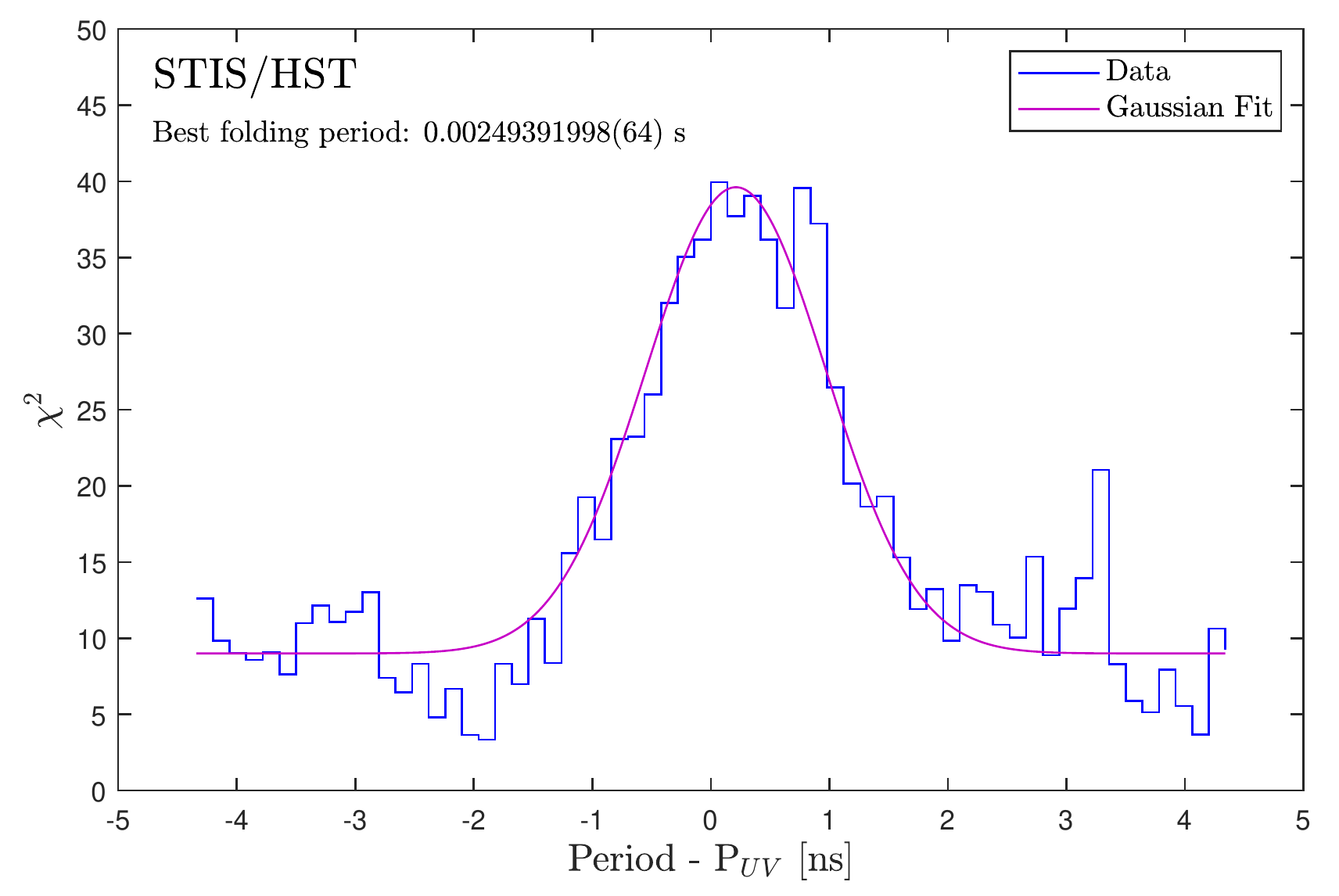}
\caption{\footnotesize Epoch folding search analysis of SAX\,J1808.4$-$3658. \emph{Top}: optical distribution of chi-squared values (in blue) as a function of trial periods. \emph{Bottom}: ultraviolet distribution of chi-squared values (in blue) as a function of trial periods.
A total number of 63 trial periods, divided into 10 phase bins, around the neutron star spin period were used for the analysis of both datasets.  
Period steps were set to 9.4$\times$10$^{-11}$\,s and 1.4$\times$10$^{-10}$\,s for the optical and ultraviolet datasets, respectively. The best-fitting Gaussian models were overplotted in light purple.}
\label{EFSs}
\end{figure}
\begin{figure}
\centering
\includegraphics[scale=0.7]{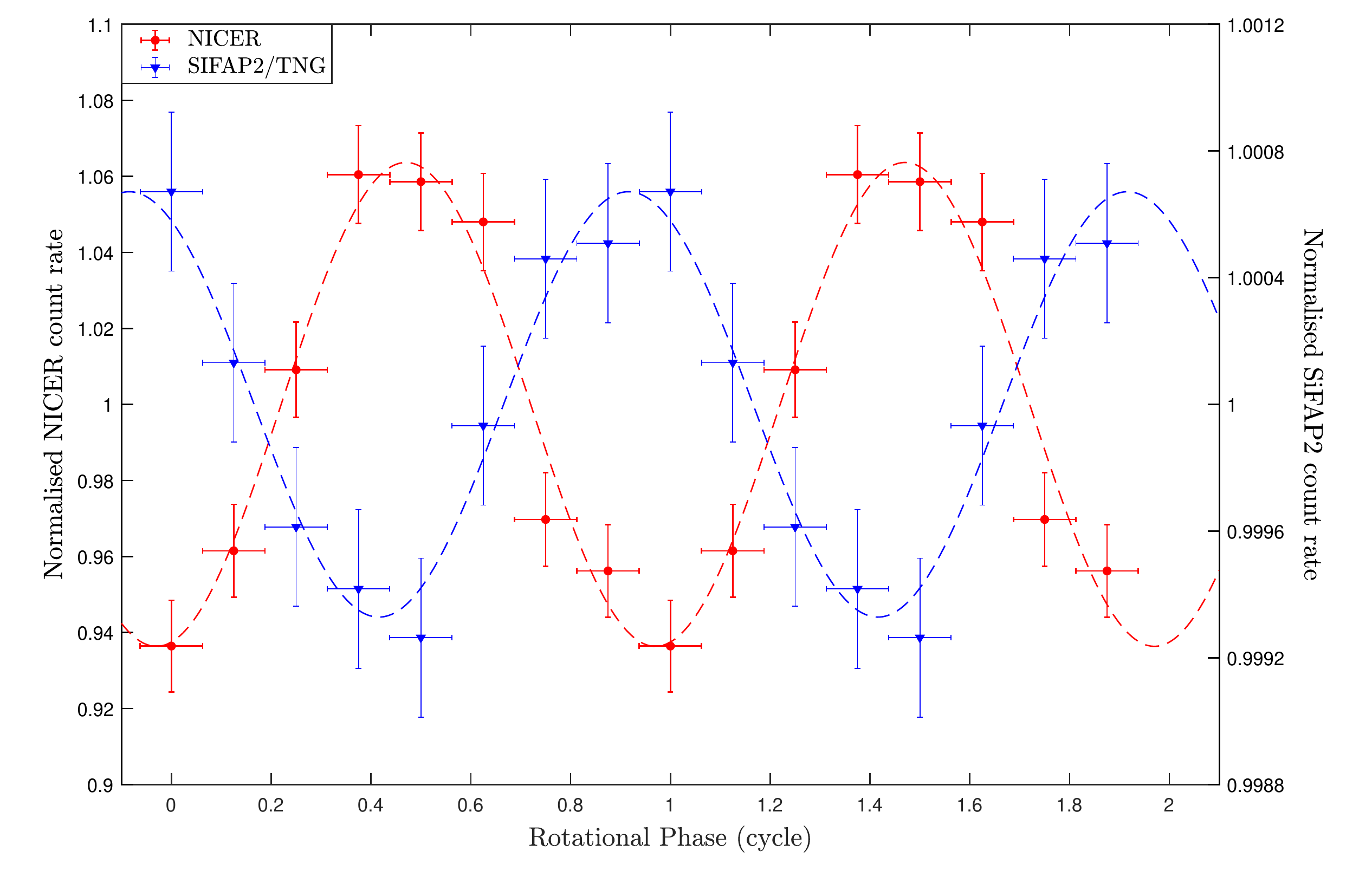}
\caption{\footnotesize Pulse profiles of SAX\,J1808.4--3658 in the X-ray (red) and optical (blue) bands (note the different scale used). Light curves were folded using the timing solution reported in Table\,\ref{Tab1}.  
The phase shift between the two curves is $\Delta\phi=(0.55\pm0.02)$, corresponding to a time lag of $\tau = (1.38\pm0.06)$\,ms.}
\label{Fig_PulseProfiles}
\end{figure}
\begin{figure}
\centering
\includegraphics[scale=0.8]{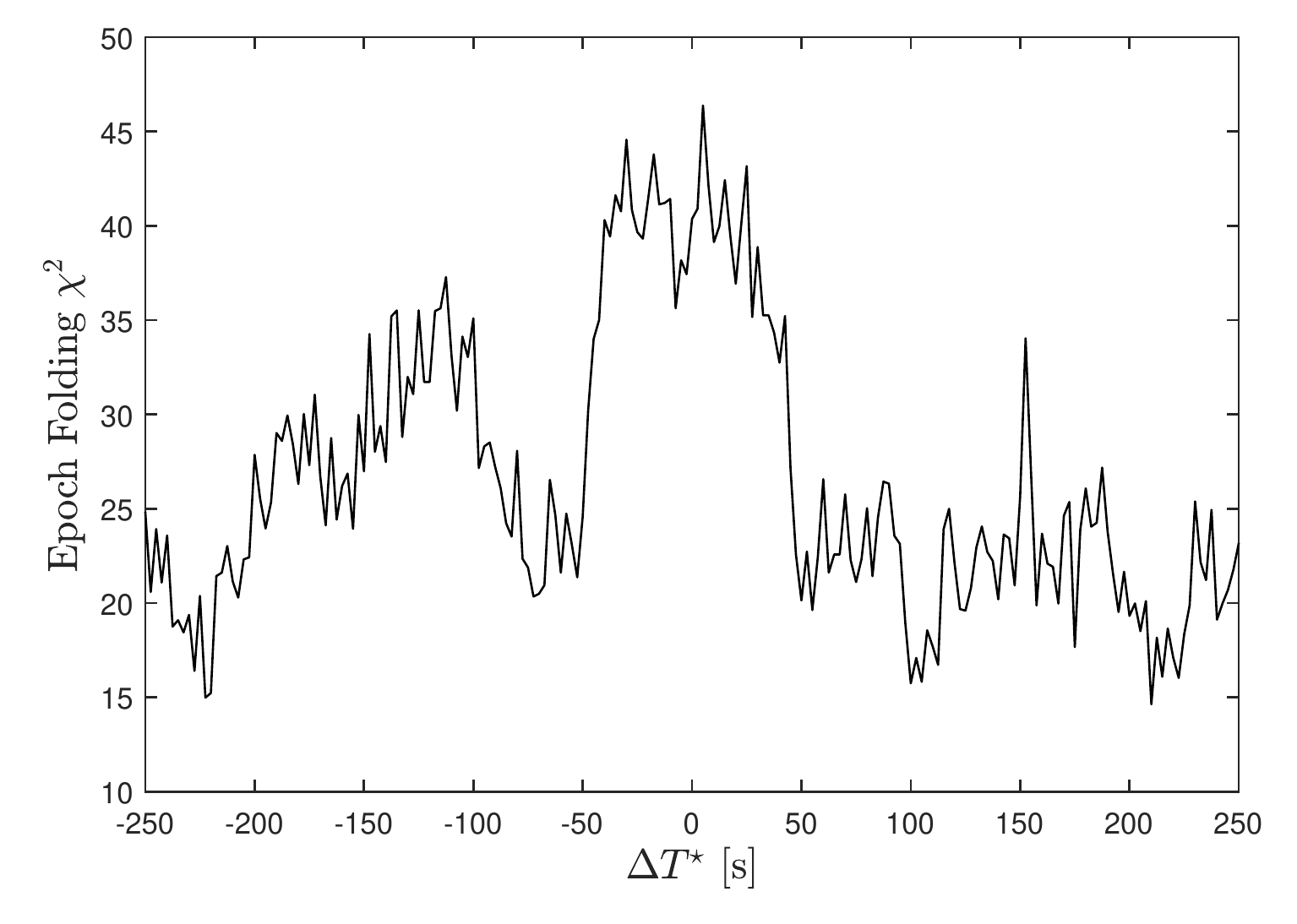}
\caption{\footnotesize Distribution of chi-squared values obtained by varying the pulsar ascending node epoch ($T^{\star}$). The distribution was obtained with an epoch folding search with $n = 8$ phase bins around the spin period expected at the epoch of the SiFAP2 observation from the X-ray analysis corrected with the value of the projection of the semi-major axis and differing by $\Delta T^{\star}$ from the best fitting value (see Table~\ref{Tab1}). The analysis covers a shift of $\pm 250$\,s around $T^{\star}$ with a time step of $2.5$\,s.}

\label{Fig_deltaTasc}
\end{figure} 

\begin{figure}
\centering
\includegraphics[scale=0.8]{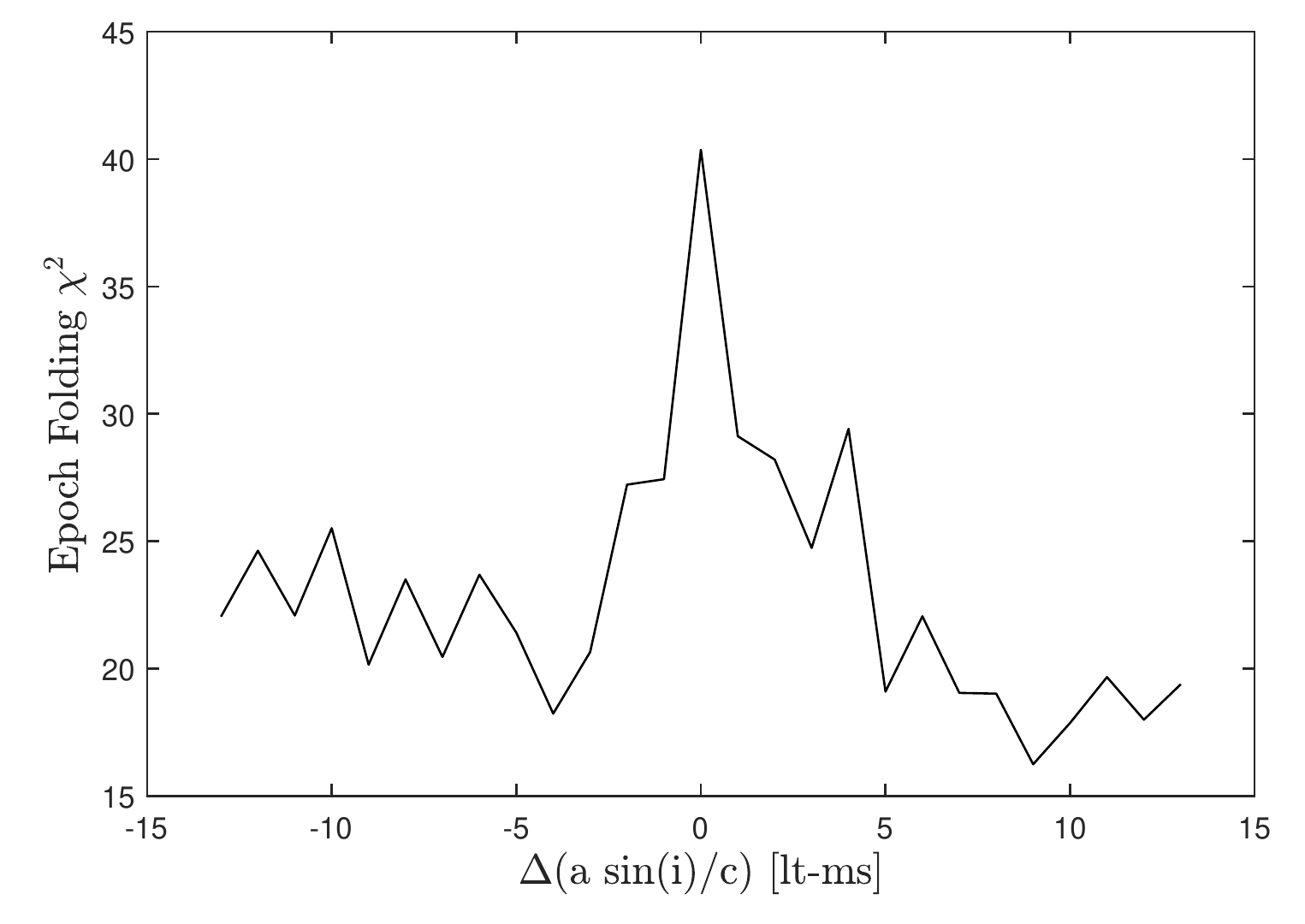}
\caption{\footnotesize Distribution of chi-squared values obtained by varying the projection of the semi-major axis of the pulsar ($a sin(i)/c$). The distribution was obtained with an epoch folding search with $n = 8$ phase bins around the spin period expected at the epoch of the SiFAP2 observation from the X-ray analysis corrected with the value of the time of ascending node and differing by $\Delta (a sin(i)/c)$ from the best fitting value (see Table~\ref{Tab1}). The analysis covers a shift of $\pm 13$\,lt-ms around $a sin(i)/c$ with a step of $1$\,lt-ms.}

\label{Fig_delta_asini}
\end{figure} 
\subsection{Spectral energy distribution.}

Fig.\,\ref{Fig9} shows the spectral energy distribution (SED) of the total and pulsed emissions in the optical, UV, and X-ray bands during the August 7, 2019 observations. The optical and UV magnitudes were corrected for interstellar extinction using the empirical relation\cite{Foight} $A_V=N_{\rm H}/(2.87 \pm 0.12) \times 10^{21}$ cm$^{-2}$, where $N_{\rm H}=2.1\times 10^{21}$\,cm$^{-2}$ is the hydrogen column density along the line of sight to SAX\,J1808.4$-$3658\cite{DiSalvo2019}. We used the extinction curves by \cite{F99} to obtain the reddening $A_{\lambda}$ in different bandpasses\cite{Schlafly}. The optical monitoring data in $B, V, R, i^{'}$ bands ($B, V, R$ in the Vega system and $i'$ in the AB system) were taken with the 2-m Faulkes Telescope South (at Siding Spring, Australia) and Las Cumbres Observatory (LCO) 1-m robotic telescopes in Chile, South Africa and Australia (see \cite{Elebert, Tudor}). The Faulkes/LCO magnitudes were calculated using the `X-ray Binary New Early Warning System' (XB-NEWS; see \cite{Russell2019, Pirbhoy} for details) data analysis pipeline.
To estimate the optical flux of SAX\,J1808.4$-$3658 at the epoch of our SiFAP2 observation, we interpolated the data of the LCO/Faulkes light curve. 
The SAX\,J1808.4$-$3658 outburst was also monitored with the Ultraviolet and Optical Telescope (UVOT\cite{Roming}) on board the {\em Neil Gehrels Swift Observatory}\cite{Gehrels} using different UV filters. The UV counterpart was detected in ten consecutive observations carried out between August 6 and September 14. The dereddened magnitude measured in the UVOT.UVM2 filter on August 7 (central wavelength of 224.6\,nm and full-width at half-maximum of 49.8\,nm) was 14.4$\pm$0.1\,mag (Vega system).
For the X-ray band, we extracted the background-subtracted spectrum from the same event file used above to evaluate the phase difference between the X-ray and optical pulses, employing the \textit{nibackgen3C50} background modelling tool available at \url{https://heasarc.gsfc.nasa.gov/docs/nicer/tools/nicer_bkg_est_tools.html} (total exposure time of $\sim2.7$\,ks). We assigned the latest versions of the NICER redistribution matrix (``nixtiref20170601v002.rmf'') and ancillary response file (``nixtiaveonaxis20170601v004.arf'') to the spectrum, and grouped it so as to contain at least 200 counts in each energy channel. The spectral analysis was performed with the Xspec package\cite{Arnaud} (version 12.11.1) and was limited to the energy band between 0.5 and 5\,keV, where the source was above the background.

\begin{figure}
\includegraphics[scale=0.34]{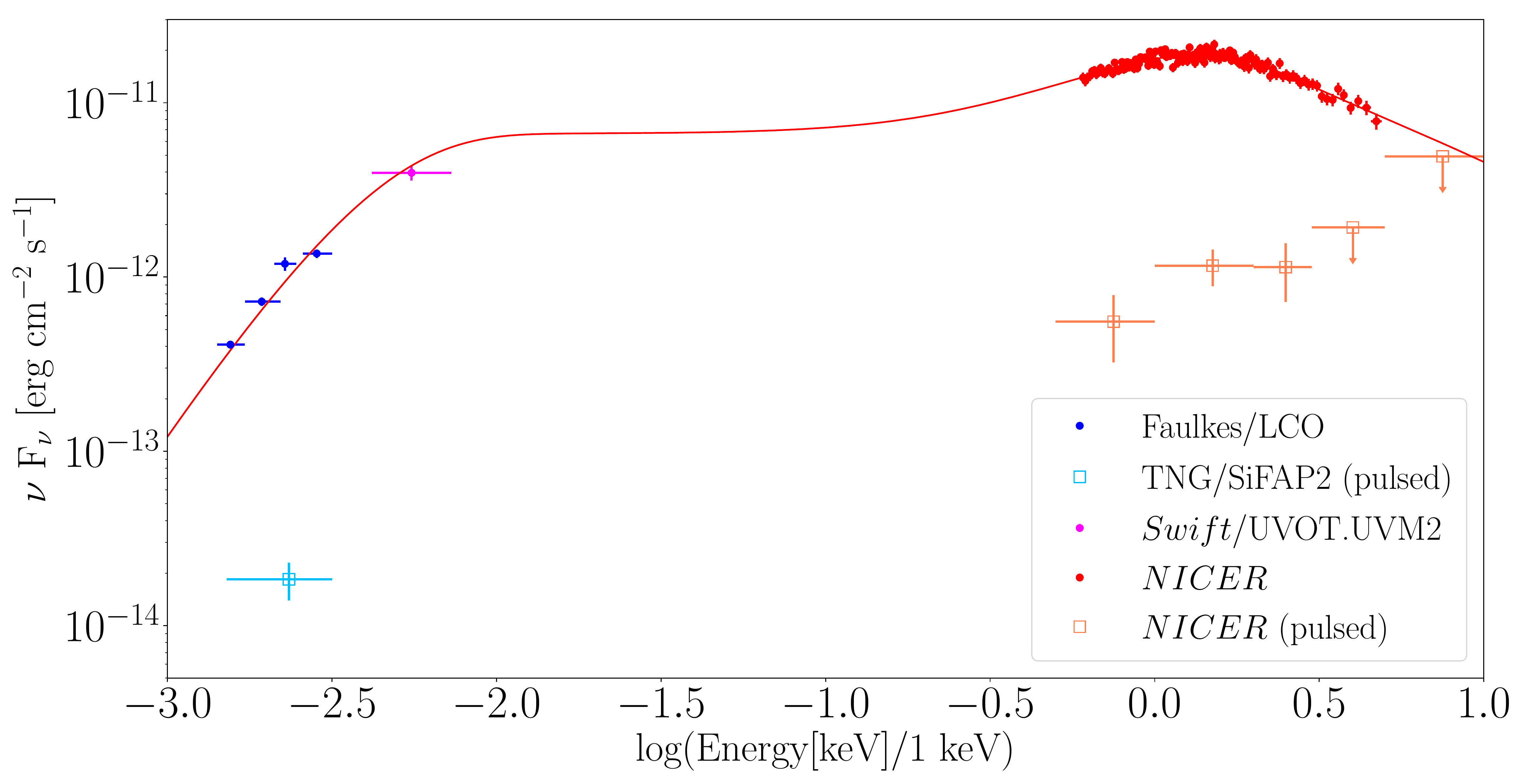}
\caption{\footnotesize Spectral energy distributions (SEDs) for the total and pulsed emissions of SAX\,J1808.4$-$3658 on August 7, 2019 corrected for interstellar extinction. The total X-ray fluxes observed by  XTI/NICER are plotted using red points. The UV/optical fluxes measured with {\em Swift}-UVOT.UVM2 and LCO/Faulkes are
shown with magenta and blue points, respectively.  For the analysis of the LCO/Faulkes data we used a new real-time data analysis pipeline, the X-ray Binary New Early Warning System\cite{Russell2019}.
The pulsed X-ray fluxes observed by XTI/NICER are computed over the 0.5--1, 1--2, 2--3, 3--5 and 5--10\,keV energy bands, and are plotted using orange squares. Upper limits are marked by arrows.
The average optical pulsed flux observed with SiFAP2/TNG is plotted with a light blue square.
The red solid line indicates the best-fitting composite model of the SED for the total emission (\textit{diskir} model in Xspec). Error bars represent 1-$\sigma$ uncertainties.
}
\label{Fig9}
\end{figure}

The SED, from optical and UV wavelengths to X-rays, is well fitted by a continuum model consisting of the sum of a blackbody, modeled by \textit{bbodyrad} in Xspec, and a Comptonisation component, modeled with \textit{nthcomp}\cite{Zycki}. 
To account for the effects of photoelectric absorption by neutral matter in the interstellar medium, we included in the spectral fit the Tuebingen--Boulder model, adopting the photoionisation cross-sections from \cite{Verner} and the chemical abundances from \cite{Wilms}. The equivalent hydrogen column density was held fixed at $N_{\rm H} = 2.1\times10^{21}$\,cm$^{-2}$ in the spectral fit\cite{Papitto2009,DiSalvo2019}. Results are listed in Table\,\ref{Tab_model1}. In this case, the best-fitting value of the blackbody temperature is $kT_{\rm BB} = 332 \pm 4$\,eV, with an estimated radius of the (spherical) emission region of $3.58 \pm 0.08$\,km for a distance to the source of 3.5\,kpc. For the \textit{nthcomp} component, we fixed the electron temperature and photon index to the best-fitting values found for the broad-band X-ray spectrum ($kT_{\rm e}$ fixed at 50\,keV and the photon index fixed at 1.9\cite{Sanna2019}, corresponding to an optical depth of about 2 for the Comptonisation region). We obtained a seed-photon temperature of $kT_{\rm seed} = 1.48 \pm 0.03$\,eV, in the hypothesis of a blackbody spectrum for the seed photons. We can therefore evaluate the radius of the emitting (spherical) region of the seed photon spectrum by using the relation given by \cite{Zand}. We find a radius $\sim 10^{10}-10^{11}$\,cm, which is compatible with the inferred size of the accretion disc in this system. This may indicate that the source of seed photons for the Comptonisation may come from the outer regions of the system, which may contribute to most of the optical/UV emission of the source.  This is in agreement with the widely accepted paradigm that most of the optical/UV emission in (black hole or neutron star) LMXB systems is produced in the outer regions of the accretion disc as the result of X-ray reprocessing\cite{Vrtilek, Paradijs, Russell}.

\begin{table*}
\caption{Spectral fit parameters of SAX\,J1808.4$-$3658 from optical and UV bands to the X-ray band using the absorbed \textit{bbodyrad} plus \textit{nthcomp} model in Xspec.}
\begin{center}
\resizebox{0.65\columnwidth}{!}{
\begin{tabular}{@{}l l}
\hline 
Parameter & Value \\
\hline
Absorption column density ($N_{\rm H}$) & $2.1\times10^{21}$\,cm$^{-2}$ (fixed) \\
Black-body temperature  ($kT_{\rm BB}$) & $332\pm 4$\,eV \\
Radius of black-body emission region ($R_{\rm BB}$)$^{(a)}$ & $3.58 \pm 0.08$\,km \\
Electron cut-off temperature ($kT_{\rm e}$)$^{(b)}$ & $50$\,keV  (fixed) \\
Asymptotic power-law photon index ($\Gamma$)$^{(c)}$ & $1.9$  (fixed) \\
Seed-photon temperature ($kT_{\rm seed}$) & $1.48 \pm 0.03$\,eV \\
Normalisation ($N_{{\rm nthComp}}$) & $(5.8\pm0.1)\times10^{-3}$\\ 
Luminosity ($L_X$)$^{(d)}$ & $6.22_{-0.04}^{+0.05}\times10^{34}$ erg\,s$^{-1}$\\
\hline
Reduced chi-squared, $\chi^2_{{\rm red}} (\mbox{d.o.f.})$ & $1.15$ (152) \\
Null-hypothesis probability & 0.097\\
\hline
\end{tabular}
}
\end{center}
\footnotesize Parameters derived from the spectral modelling of data acquired by NICER/XTI, Faulkes/LCO and {\em Swift}-UVOT 
on August 7, 2019 close to the epoch of the SiFAP2 observations. Uncertainties on each parameter are 
quoted at a confidence level of 1$\sigma$.
\newline $^{(a)}$ The radius of the black-body emission region was evaluated assuming a source 
distance\cite{Galloway} of $3.5$\,kpc.
\newline$^{(b-c)}$ The narrow energy band adopted for the spectral modelling (0.5--5\,keV) does not allow tight constraints on the parameters of the 
Comptonised component. Therefore, in the spectral fits, the electron cut-off temperature was held fixed  to $kT_{{\rm e}}=50$\,keV, a value comparable to those typically measured in the broad-band spectra  of AMXPs (including SAX\,J1808.4$-$3658\cite{DiSalvo2019}); the asymptotic power-law photon index was fixed to  $\Gamma=1.9$, that is, the value measured using an observation with the {\em NuSTAR} satellite on 10--11 August\cite{Sanna2019}.
\newline $^{(d)}$ X-ray luminosity. It was evaluated over the 0.5--10\,keV range using the convolution model \textit{cflux} in Xspec, assuming a distance\cite{Galloway} of 3.5\,kpc and isotropic emission.
\label{Tab_model1}
\end{table*}

In order to obtain a more physical interpretation of the broad-band emission of SAX\,J1808.4$-$3659, we fitted the SED using a model that accounts for the emission of a truncated accretion disc irradiated by a hot Comptonising accretion flow (\textit{diskir}\cite{Gierlinski} in the Xspec notation). In this model, the X-ray emission consists of thermal radiation from the disc and a hard tail produced by Comptonisation of soft seed photons in a hot plasma of energetic electrons. A fraction of this radiation is intercepted by the outer regions of the disc, reprocessed and re-emitted 
in the optical and UV bands. 
This model often describes well the broad-band emission of bright LMXBs. 
The column density was again fixed to $N_{\rm H} = 2.1\times10^{21}$\,cm$^{-2}$ in the spectral fit.
We obtained a statistically acceptable description of the data, with a reduced chi-squared of $\chi^2_{{\rm red}} = 0.89$ for 152 degrees of freedom (d.o.f.). Results are listed in Table\,\ref{Tab2}. 
According to this model, the inner disc has an intrinsic (i.e., not irradiated) temperature of $\sim$ 200\,eV and a radius of $\sim 25$\,km (inferred from the model normalisation and assuming an inclination angle of $\sim$50$^{\circ}$, as derived from modeling of the multi-band light
curve in quiescence using a Markov chain Monte Carlo technique\cite{Wang}). The inner disc provides the seed photons Comptonised by the hot electron cloud close to the neutron star (possibly the accretion column) producing the main Comptonisation continuum observed in the X-ray band; the electron temperature of this component has been fixed to the value found from modelling of the X-ray spectrum\cite{DiSalvo2019}, while the photon index was allowed to vary (best-fitting value of $\Gamma = 2.76\pm0.06$). A fraction $7.1_{-1.2}^{+1.5} \%$ of this hard flux is reprocessed in the outer disc, whose radius is about $10^4$ times larger than the inner disc radius, $R_{\rm out} \sim 10^{10}$\,cm. This is comparable to the size of the Roche lobe of the neutron star and thus the size of the disc.  

In order to investigate the origin of the pulsed emission in the optical/UV band, we also extracted the pulsed SED on August 7--8, 2019. For the X-ray band, we firstly evaluated the values for the background-subtracted pulse rms amplitudes over the energy ranges 0.5--1, 1--2 and 2--3\,keV, as well as the 3$\sigma$ upper limits over the ranges 3--5 and 5--10\,keV. We calculated the de-absorbed X-ray fluxes over these same energy ranges by extrapolating the best-fitting model for the total emission (pulsed plus unpulsed), and multiplied them by the corresponding values (or upper limits) of the pulse rms amplitude to obtain integrated pulsed fluxes. We then
multiplied the values so evaluated for the ratio between the mid-point energy of the interval and the width of the energy interval to derive pulsed fluxes (and upper limits) in $\nu$\,$F_{\nu}$ units. To convert the pulsed optical fluxes into $\nu$\,$F_{\nu}$ units, we multiplied dereddened fluxes by the filter full-width at half-maximum (89\,nm, 84\,nm, 158\,nm and 154\,nm for the $B$, $V$, $R$, and $i'$ filters, respectively; we converted the AB magnitude of filter $i'$ in the Vega system). We then co-added the fluxes in the four different bands to obtain one single value for the pulsed flux covering the SiFAP2 operating band (320--900\,nm) and multiplied such value by the optical pulse fractional amplitude. A total of 4 SED data points were obtained in this way (see Fig.~\ref{Fig9}), not enough to perform a meaningful modeling. We could test just two-parameters models on these data, such as blackbody, thermal emission from the disc, bremsstrahlung, and power-law. A power-law model fit to the SED data points for the pulsed optical and X-ray emissions yields a reduced chi-squared value of $\chi^2_{{\rm red}} = 0.4$ for 2 d.o.f, and a functional dependence of the form $\nu$\,$F_{\nu}$\,$\propto$\,$\nu^{(0.62\pm0.04)}$. 



\begin{table*}
\caption{Spectral fit parameters of SAX\,J1808.4$-$3658 from optical and UV bands to the X-ray band using the absorbed \textit{diskir} model in Xspec.}
\begin{center}
\resizebox{0.6\columnwidth}{!}{
\begin{tabular}{@{}l l}
\hline 
Parameter & Value \\
\hline
Absorption column density ($N_{\rm H}$) & $2.1\times10^{21}$\,cm$^{-2}$ (fixed) \\
Temperature of the unilluminated inner disc ($kT_{{\rm in}}$) & $191_{-21}^{+22}$\,eV \\
Asymptotic power-law photon index ($\Gamma$) & $2.76\pm0.06$ \\
Electron cut-off temperature ($kT_{\rm e})^{(a)}$ & $50$\,keV  (fixed) \\
Luminosity ratio ($L_{{\rm Compt}}/L_{{\rm disc}})^{(b)}$ & $4.2_{-1.2}^{+2.1}$ \\
Fraction of reprocessed emission ($f_{{\rm in}})^{(c)}$ & 10\,\% (fixed) \\
Fraction of reprocessed emission ($f_{{\rm out}})^{(c)}$ & $7.1_{-1.2}^{+1.5}$\,\%  \\
Inner disc radius ($R_{\rm in}\sqrt{\cos{i}}$)$^{(d)}$ & $21.9_{-2.2}^{+2.3}$\,km \\
Inner radius of illuminated disc ($R_{{\rm irr}})^{(e)}$ & 1.1\,$R_{\rm in}$ (fixed)  \\
Outer disc radius (${\rm {Log}}[R_{{\rm out}}/R_{{\rm in}}])^{(f)}$ & $4.19\pm0.06$ \\
Luminosity ($L_X$)$^{(g)}$ & $5.68_{-0.04}^{+0.05}\times10^{34}$ erg\,s$^{-1}$\\
\hline
Reduced chi-squared, $\chi^2_{{\rm red}} (\mbox{d.o.f.})$ & $0.891$ (152) \\
Null-hypothesis probability & 0.828\\
\hline
\end{tabular}
}
\end{center}
\scriptsize Parameters derived from the spectral modelling of data acquired by NICER/XTI, Faulkes/LCO and {\em Swift}-UVOT 
on August 7, 2019 close to the epoch of the SiFAP2 observations. Uncertainties on each parameter are 
quoted at a confidence level of 1$\sigma$, whereas upper limits are reported at a confidence level of 3$\sigma$.
\newline$^{(a)}$ The narrow energy band adopted for the spectral modelling (0.5--5\,keV) does not allow tight constraints on this parameter. Therefore, in the spectral fits, it was held fixed to $kT_{{\rm e}}=50$\,keV, a value comparable to those typically measured in the broad-band spectra 
of AMXPs (including SAX\,J1808.4$-$3658\cite{DiSalvo2019}).
\newline$^{(b)}$ Ratio between the luminosity of the Comptonised component and that of the non-illuminated disc.
\newline$^{(c)}$ Fraction of bolometric flux which is intercepted and reprocessed in the inner ($f_{{\rm in}}$) and outer ($f_{{\rm out}}$) disc. The former was fixed at 10\,\% in the spectral fits. 
\newline$^{(d)}$ Inner radius of the accretion disc. It was evaluated assuming a source distance\cite{Galloway} of $3.5$\,kpc and adopting a color correction factor\cite{Kubota}. $i$ is the disc inclination.
\newline$^{(e)}$ Radius of the inner disc region that is illuminated by the corona. It was fixed at 1.1\,$R_{\rm in}$ in the spectral fits.
\newline$^{(f)}$ Ratio between the outer and the inner disc radii.
\newline $^{(g)}$ X-ray luminosity. It was evaluated over the 0.5--10\,keV range using the convolution model \textit{cflux} in Xspec, assuming a distance\cite{Galloway} of 3.5\,kpc and isotropic emission.
\label{Tab2}
\end{table*}

\subsection{Models.}

The fundamental photon cyclotron energy emitted by electrons in the magnetic field of SAX\,J1808.4$-$3658 at the base of the accretion column ($B\sim 3.5 \times 10^8$\,G \cite{DiSalvo2003, Burderi2006, Sanna2017}) is $E_{\rm cyc}$ = 11.6 ($B/10^{12}$\,G)\,keV $\sim 4.1 (r/R_{NS})^{-3}$\,eV. At these energies, the UV and optical pulsed luminosity produced by optically thick cyclotron emission can be estimated as\cite{Thompson} $L_{cyc} =A \int_{\nu_1}^{\nu_2} (2 \pi k T_e \nu^2/ 3 c^2) d\nu$, where $\nu_1$ and $\nu_2$ are the boundary frequencies in the SiFAP2 or STIS/G230L energy range and $kT_e$ is the electron temperature. $A\sim~\pi R_{\rm NS}^2 (R_{\rm NS}/r_c)$ is the hot-spot area of the accreting polar cap\cite{Frank2002}, where $R_{\rm NS}$ = 10\,km is the neutron star radius and $r_c$ is the corotation radius (the distance at which the Keplerian velocity of matter in the disc equals the velocity of the neutron star magnetosphere). The area of the accreting region is $\sim$100\,km$^2$ at most. The maximum cyclotron luminosity (using\cite{Poutanen2003} $kT_e$ = 100\,keV) from electrons in the accretion column is: $L_{cyc(opt)}\sim~10^{29}$\,erg\,s$^{-1}$ in the 320-900\,nm band and $L_{cyc(UV)} \sim 6\times 10^{29}$\,erg\,s$^{-1}$ in the 165-310\,nm band. These values are orders of magnitude smaller than the observed UV and optical pulsed luminosity, and do not favor a cyclotron emission scenario for the UV and optical bands.

An alternative scenario involves the presence of the accretion disc which does not prevent the rotation-powered mechanism from working. In this way, optical and UV pulsations could be produced by a rotation-driven pulsar, as in the isolated neutron stars\cite{Romani, Torres, Torres2019}. 
The long-term spin-down energy loss rate can be estimated as $\dot{E}_{sd}=4 \pi^2 I \nu \dot{\nu} \sim 2 \times 10^{34}$\,erg\,s$^{-1}$, where $I$ is the moment of inertia of the neutron star, $\nu$ is the neutron star spin frequency (see Tab.~\ref{Tab1}) and $\dot{\nu}$ is the first time derivative of the neutron star spin frequency\cite{Bult}. Comparing this value with the observed pulsed luminosity leads to the magneto-rotational efficiencies for SAX\,J1808.4$-$3658 of $\eta_{UV}~\sim 1 \times 10^{-2}$ and $\eta_{opt}\sim 6 \times 10^{-4}$ in the $B$ band with $\eta_{UV/opt}=L_{pulsed(UV/opt)}/ \dot{E}_{sd}$ computed at the time of our observations. Optical and UV pulses have been observed only in five rotation-powered pulsars\cite{Mignani, Mignani2019}, all isolated, slower-rotating, younger and with high magnetic fields ($> 10^{12}$\,G). The efficiency in converting spin-down power to UV/optical luminosity was determined in PSR B0540$-$69 and in the Crab pulsar\cite{Ambrosino, Mignani}. Our estimates are orders of magnitude higher than the Crab pulsar efficiencies of $\eta_{UV}~\sim 2 \times 10^{-5}$ in the 165--310\,nm band and $\eta_{opt}~\sim 5 \times 10^{-6}$ in the $B$ band. These high efficiencies in SAX\,J1808.4$-$3658 would call for the existence of a (so far unknown) physical process that is able to enhance the spin-down powered emission in the presence of an accretion disc.

\subsection{Comparison between SAX J1808.4--3658 and PSR J1023+0038.}

Before SAX\,J1808.4$-$3658, fast optical pulsations were detected only from the transitional millisecond pulsar PSR\,J1023$+$0038\cite{Ambrosino} while it was lingering in
an X-ray sub-luminous disc state at an average (0.5--10\,keV) X-ray luminosity of $L_X \sim 4 \times 10^{33}$\,erg\,s$^{-1}$. 
The pulsed optical luminosity was high also in that case ($\approx 10^{31}$\,erg\,s$^{-1}$), considering that the source was releasing a spin-down power of $4.3\times 10^{34}$\,erg\,s$^{-1}$ in the radio pulsar state. Optical and X-ray pulses were almost phase aligned and detected only during the so-called {\it high} X-ray luminosity mode in which the source spends $\sim70\%$ of the time, but were seen to suddenly disappear in the lower luminosity mode, suggesting a common underlying process\cite{Papitto2019}. Both the optical and X-ray pulsations observed from PSR\,J1023$+$0038 are thought to originate from synchrotron radiation in the intrabinary shock just beyond the light cylinder radius, where the wind of relativistic particles ejected by the pulsar meets the accretion disc\cite{Papitto2019, Veledina, Campana2019}. 
Optical pulsations from SAX\,J1808.4$-$3658 were detected in an 
intermediate stage of the outburst, when the X-ray luminosity had not yet peaked. Yet, the corresponding X-ray luminosity exceeded that of PSR\,J1023+0038 by about an order of magnitude ($L_X \sim 6 \times 10^{34}$\,erg\,s$^{-1}$). Optical pulses lagged the X-ray ones by $\sim 1.4$\,ms, that is, they were almost in anti-phase. During this outburst and the previous ones, the X-ray spectral and timing properties of SAX\,J1808.4$-$3658 did not show any evidence for transitions to a non-accreting regime. These arguments suggest that its X-ray, UV and optical pulses can be hardly explained by invoking the same physical mechanism.

\end{methods}


\begin{thebibliography}{1}


\bibitem{Alpar} Alpar, M.~A., Cheng, A.~F., Ruderman, M.~A. \& Shaham, J.\ A new class of radio pulsars.\ \textit{Nature} \textbf{300}, 728--730 (1982).

\bibitem{Wijnands} Wijnands, R. \& van der Klis, M.\ A millisecond pulsar in an X-ray binary system.\
\textit{Nature} \textbf{394}, 344--346 (1998).

\bibitem{Campana2018} Campana, S. \& Di Salvo, T.\ Accreting Pulsars: Mixing-up Accretion Phases in Transitional Systems.\ \textit{Astrophysics and Space Science Library} \textbf{457}, 149--184 (2018).

\bibitem{Bult} Bult, P. {\it et al.}\ Timing the Pulsations of the Accreting Millisecond Pulsar SAX\,J1808.4$-$3658 during Its 2019 Outburst. \textit{Astrophys. J.} \textbf{898}, 38 (2020).

\bibitem{Torres} Torres, D.~F. Order parameters for the high-energy spectra of pulsars.\ \textit{Nat. Astron.} \textbf{2}, 247--256 (2018). 

\bibitem{Harding2018} Harding, A.~K., Kalapotharakos, C., Barnard, M. \& Venter, C.\ Multi-TeV Emission from the Vela Pulsar.\ \textit{Astrophys. J. Lett.} \textbf{869}, L18 (2018).

\bibitem{Chakrabarty} Chakrabarty, D. \& Morgan, E.~H.\ The two-hour orbit of a binary millisecond X-ray pulsar.\ \textit{Nature} \textbf{394}, 346--348 (1998).

\bibitem{Archibald} Archibald, A.~M. {\it et al.}\ A Radio Pulsar/X-ray Binary Link.\ \textit{Science} \textbf{324}, 1411--1414 (2009). 

\bibitem{Papitto2013} Papitto, A. {\it et al.} \ Swings between rotation and accretion power in a binary millisecond pulsar.\ \textit{Nature} \textbf{501}, 517–-520 (2013).

\bibitem{Watts} Watts, A.~L. {\it et al.} \ Colloquium: Measuring the neutron star equation of state using x-ray timing.\ \textit{Reviews of Modern Physics} \textbf{88}, 021001 (2016).

\bibitem{Wijnands2006} Wijnands, R.\ Accretion-Driven Millisecond X-ray Pulsars.\ \textit{Trends in Pulsar Research} \textbf{53},  (2006).

\bibitem{Patruno} Patruno, A. \& Watts, A.~L.\ Accreting Millisecond X-ray Pulsars. \ \textit{Astrophysics and Space Science Library} \textbf{143}, 143--208 (2021).

\bibitem{Liu} Liu, Q.~Z., van Paradijs, J., \& van den Heuvel, E.~P.~J. \ A catalogue of low-mass X-ray binaries in the Galaxy, LMC, and SMC (Fourth edition). \ \textit{\aap} \textbf{469}, 807--810 (2007).

\bibitem{Ambrosino} Ambrosino, F. \textit{et al.} \ Optical pulsations from a transitional millisecond pulsar. \ \textit{Nat. Astron.} \textbf{1}, 854-858 (2017).

\bibitem{Papitto2019} Papitto, A. \textit{et al.} \ Pulsating in Unison at Optical and X-Ray Energies: Simultaneous High Time Resolution Observations of the Transitional Millisecond Pulsar PSR\,J1023$+$0038. \ \textit{\apj} \textbf{882}, 104 (2019).

\bibitem{Veledina} Veledina, A., N{\"a}ttil{\"a}, J., \& Beloborodov, A.~M.\ Pulsar Wind-heated Accretion Disk and the Origin of Modes in Transitional Millisecond Pulsar PSR\,J1023$+$0038. \ \textit{\apj} \textbf{884}, 144 (2019).

\bibitem{Campana2019} Campana, S. \textit{et al.}\ Probing X-ray emission in different modes of PSR\,J1023$+$0038 with a radio pulsar scenario. \ \textit{\aap} \textbf{629}, L8 (2019).

\bibitem{Wang} Wang, Z. \textit{et al.} \ Multiband Studies of the Optical Periodic Modulation in the X-Ray Binary SAX\,J1808.4$-$3658 during Its Quiescence and 2008 Outburst. \ \textit{\apj} \textbf{765}, 151 (2013).

\bibitem{Galloway} Galloway, D.~K. \& Cumming, A.\ Helium-rich Thermonuclear Bursts and the Distance to the Accretion-powered Millisecond Pulsar SAX\,J1808.4$-$3658. \ \textit{\apj} \textbf{652}, 559--568 (2006).

\bibitem{Gilfanov} Gilfanov, M. \textit{et al.} \ The millisecond X-ray pulsar/burster SAX\,J1808.4$-$3658: the outburst light curve and the power law spectrum. \ \textit{\aap} \textbf{338}, L83--L86 (1998).

\bibitem{Stella2k} Stella, L., Campana, S., Mereghetti, S., Ricci, D. \& Israel, G.~L. \ The Discovery of Quiescent X-Ray Emission from SAX\,J1808.4$-$3658, the Transient 2.5 Millisecond Pulsar. \ \textit{\apjl} \textbf{537}, L115-L118 (2000).

\bibitem{Giles} Giles, A.~B., Hill, K.~M., \& Greenhill, J.~G.\ The optical counterpart of SAX\,J1808.4$-$3658, the transient bursting millisecond X-ray pulsar. \ \textit{\mnras} \textbf{304}, 47--51 (1999).

\bibitem{Gierlinski} Gierli{\'n}ski, M., Done, C., \& Barret, D.\ Phase-resolved X-ray spectroscopy of the millisecond pulsar SAX\,J1808.4$-$3658. \ \textit{\mnras} \textbf{331}, 141--153 (2002).

\bibitem{Hartman2008} Hartman, J.~M. \textit{et al.} \ The Long-Term Evolution of the Spin, Pulse Shape, and Orbit of the Accretion-powered Millisecond Pulsar SAX\,J1808.4$-$3658. \ \textit{\apj} \textbf{675}, 1468--1486 (2008).

\bibitem{Burderi2006} Burderi, L. \textit{et al.} \ Order in the Chaos: Spin-up and Spin-down during the 2002 Outburst of SAX\,J1808.4$-$3658. \ \textit{\apjl} \textbf{653}, L133--L136 (2006).

\bibitem{Thompson} Thompson, A.~M. \& Cawthorne, T.~V.\ Cyclotron emission from white dwarf accretion columns. \ \textit{\mnras} \textbf{224}, 425--434 (1987).

\bibitem{Basko} Basko, M.~M. \& Sunyaev, R.~A.\ Radiative transfer in a strong magnetic field and accreting X-ray pulsars.\ \textit{\aap} \textbf{42}, 311–-321 (1975).

\bibitem{Melrose2017} Melrose, D.~B.\ Coherent emission mechanisms in astrophysical plasmas.\ \textit{Reviews of Modern Plasma Physics} \textbf{1}, 5 (2017). 

\bibitem{Pacini} Pacini, F. \& Salvati, M.\ The optical luminosity of very fast pulsars.\ \textit{Astrophys. J.} \textbf{274}, 369–-371 (1983). 

\bibitem{Mignani} Mignani, R.~P.\ Optical, ultraviolet, and infrared observations of isolated neutron stars.\ \textit{Adv. Space. Res.} \textbf{47}, 1281--1293 (2011).


\bibitem{Parfrey2017_2} Parfrey, K. \& Tchekhovskoy, A.\ General-relativistic Simulations of Four States of Accretion onto Millisecond Pulsars.\ \textit{Astrophys. J. Lett.} \textbf{851}, L34 (2017).


\bibitem{Parfrey2016} Parfrey, K., Spitkovsky, A. \& Beloborodov, A.~M.\ Torque Enhancement, Spin Equilibrium, and Jet Power from Disk-Induced Opening of Pulsar Magnetic Fields.\ \textit{Astrophys. J.} \textbf{822}, 33 (2016).

\bibitem{Kluziniak2007} Klu{\'z}niak, W. \& Rappaport, S.\ Magnetically Torqued Thin Accretion Disks.\ \textit{Astrophys. J.} \textbf{671}, 1990–2005 (2007).

\bibitem{Coroniti} Coroniti, F.~V.\ Magnetically Striped Relativistic Magnetohydrodynamic Winds: The Crab Nebula Revisited.\ \textit{Astrophys. J.} \textbf{349}, 538 (1990).

\bibitem{Kirk2002} Kirk, J.~G., Skj{\ae}raasen, O. \& Gallant, Y.~A.\ Pulsed radiation from neutron 
star winds.\ \textit{Astron. Astrophys. Lett.} \textbf{388}, L29--L32 (2002).

\bibitem{Rybicki} Rybicki, G.~B. \& Lightman, A.~P.\ \textit{Radiative Processes In Astrophysics.}\ A Wiley-Interscience Publication, New York: Wiley (1979).

\bibitem{Ghedina} Ghedina, A. \textit{et al.} \ SiFAP2: a new versatile configuration at the TNG for the MPPC based photometer. \ \textit{Proc. SPIE} \textbf{10702}, 107025Q (2018).

\bibitem{Meddi} Meddi, F. \textit{et al.} \ A New Fast Silicon Photomultiplier Photometer. \ \textit{Publ. Astron. Soc. Pac.} \textbf{124}, 448--453 (2012).

\bibitem{Ambrosino2016} Ambrosino, F. \textit{et al.} \ The Latest Version of SiFAP: Beyond Microsecond Time Scale Photometry of Variable Objects. \ \textit{J. Astron. Instrum.} \textbf{5}, 1650005--1267 (2016).

\bibitem{Gendreau2016} Gendreau, K.~C. \textit{et al.}\ The Neutron star Interior Composition Explorer (NICER): design and development.\ \textit{Proc. SPIE} \textbf{9905}, 99051H (2016).

\bibitem{Gendreau2017} Gendreau, K. \& Arzoumanian, Z. \ Searching for a pulse. \ \textit{Nat. Astron.} \textbf{1}, 895--895 (2017).

\bibitem{Papitto2011} Papitto, A. \textit{et al.} \ Spin down during quiescence of the fastest known accretion-powered pulsar. \ \textit{\aap} \textbf{528}, A55 (2011).

\bibitem{Leahy83} Leahy, D.~A., Elsner, R.~F. \& Weisskopf, M.~C.\ On searches for periodic pulsed emission - The Rayleigh test compared to epoch folding. \ \textit{\apj} \textbf{272}, 256--258 (1983).

\bibitem{Leahy87} Leahy, D.~A.\ Searches for pulsed emission - Improved determination of period and amplitude from epoch folding for sinusoidal signals. \ \textit{\aap} \textbf{180}, 275--277 (1987).

\bibitem{Buccheri} Buccheri, R. \textit{et al.} \ Search for pulsed $\gamma$-ray emission from radio pulsars in the COS-B data. \ \textit{\aap} \textbf{128}, 245--251 (1983).

\bibitem{Foight} Foight, D.~R., G{\"u}ver, T., {\"O}zel, F. \& Slane, P.~O.\ Probing X-Ray Absorption and Optical Extinction in the Interstellar Medium Using Chandra Observations of Supernova Remnants. \ \textit{\apj} \textbf{826}, 66 (2016).

\bibitem{DiSalvo2019} Di Salvo, T. \textit{et al.} \ NuSTAR and XMM-Newton broad-band spectrum of SAX J1808.4-3658 during its latest outburst in 2015. \ \textit{\mnras} \textbf{483}, 767--779 (2019).

\bibitem{F99} Fitzpatrick, E.~L.\ Correcting for the Effects of Interstellar Extinction. \ \textit{Publ. Astron. Soc. Pac.} \textbf{111}, 63--75 (1999).

\bibitem{Schlafly} Schlafly, E.~F. \& Finkbeiner, D.~P.\ Measuring Reddening with Sloan Digital Sky Survey Stellar Spectra and Recalibrating SFD. \ \textit{\apj} \textbf{737}, 103 (2011).

\bibitem{Elebert} Elebert, P \textit{et al.} \ Optical spectroscopy and photometry of SAX\,J1808.4$-$3658 in outburst. \ \textit{\mnras} \textbf{395}, 884--894 (2009).

\bibitem{Tudor} Tudor, V. \textit{et al.} \ Disc-jet coupling in low-luminosity accreting neutron stars. \ \textit{\mnras} \textbf{470}, 324--339 (2017).

\bibitem{Russell2019} Russell, D.~M. \textit{et al.} \ Optical precursors to X-ray binary outbursts. \ \textit{Astronomische Nachrichten} \textbf{340}, 278--283 (2019).

\bibitem{Pirbhoy} Pirbhoy, S.~F. \textit{et al.} \ XB-NEWS detection of a new outburst of MAXI\,J1348$-$630. \ \textit{Astronomer's Telegram} \textbf{13451}, 1 (2020).

\bibitem{Roming} Roming, P.~W.~A. \textit{et al.} \ The Swift Ultra-Violet/Optical Telescope. \ \textit{Space Sci. Rev.} \textbf{120}, 95--142 (2005).

\bibitem{Gehrels} Gehrels, N. \textit{et al.} \ The Swift Gamma-Ray Burst Mission. \ \textit{\apj} \textbf{611}, 1005--1020 (2004).

\bibitem{Arnaud} Arnaud, K.~A.\ XSPEC: The First Ten Years. In Jacoby, G.~H. \& Barnes, J. (eds.) \ \textit{Astronomical Data Analysis Software and Systems V} \textbf{101} of \textit{Astronomical Society of the Pacific Conference Series}, 17 (1996).

\bibitem{Zycki} {\.Z}ycki, P.~T., Done, C. \& Smith, D.~A.\ The 1989 May outburst of the soft X-ray transient GS\,2023$+$338 (V404 Cyg). \ \textit{\mnras} \textbf{309}, 561--575 (1999).

\bibitem{Verner} Verner, D.~A., Ferland, G.~J., Korista, K.~T. \& Yakovlev, D.~G.\ Atomic Data for Astrophysics. II. New Analytic FITS for Photoionization Cross Sections of Atoms and Ions. \ \textit{\apj} \textbf{465}, 487 (1996).

\bibitem{Wilms} Wilms, J., Allen, A., \& McCray, R.\ On the Absorption of X-Rays in the Interstellar Medium. \  \textit{\apj} \textbf{542}, 914--924 (2000).

\bibitem{Papitto2009} Papitto, A. \textit{et al.} \ XMM-Newton detects a relativistically broadened iron line in the spectrum of the ms X-ray pulsar SAX\,J1808.4$-$3658. \ \textit{\aap} \textbf{493}, L39-L43 (2009).

\bibitem{Sanna2019} Sanna, A. \textit{et al.} \ NuSTAR observation of the latest outburst of SAX\,J1808.4$-$3658. \ \textit{Astronomer's Telegram} \textbf{13022}, 1 (2019).

\bibitem{Zand} in 't Zand, J.~J.~M. \textit{et al.}\ Discovery of the X-ray transient SAX\,J1808.4$-$3658, a likely low-mass X-ray binary.\ \textit{Astron. Astrophys. Lett.} \textbf{331}, L25--L28 (1998).

\bibitem{Vrtilek} Vrtilek, S.~D. \textit{et al.}\ Observations of Cygnus X$-$2 with IUE : ultraviolet results from a multiwavelength campaign.\ \textit{Astron. Astrophys.} \textbf{235}, 162 (1990).

\bibitem{Paradijs} van Paradijs, J. \& McClintock, J.~E.\ Optical and ultraviolet observations of X-ray binaries.\ \textit{X-ray Binaries}, 58--125 (1995).

\bibitem{Russell} Russell, D.~M. \textit{et al.}\ Global optical/infrared-X-ray correlations in X-ray binaries: quantifying disc and jet contributions. \ \textit{\mnras} \textbf{371}, 1334--1350 (2006).

\bibitem{Kubota} Kubota, A. \textit{et al.} \ Evidence for a Black Hole in the X-Ray Transient GRS\,1009$-$45. \ \textit{Publ. of the Astron. Soc. of Jpn.} \textbf{50}, 667--673 (1998).

\bibitem{DiSalvo2003} Di Salvo, T. \& Burderi, L.\ Constraints on the neutron star magnetic field of the two X-ray transients SAX\,J1808.4$-$3658 and Aql\,X$-$1. \ \textit{\aap} \textbf{397}, 723--727 (2003).

\bibitem{Sanna2017} Sanna, A. \textit{et al.} \ On the timing properties of SAX\,J1808.4$-$3658 during its 2015 outburst. \ \textit{\mnras} \textbf{471}, 463--477 (2017).

\bibitem{Frank2002} Frank, J., King, A. \& Raine, D.~J.\ \textit{Accretion Power in Astrophysics.} Cambridge, UK: Cambridge University Press, 398 (2002).

\bibitem{Poutanen2003} Poutanen, J. \& Gierli{\'n}ski, M.\ On the nature of the X-ray emission from the accreting millisecond pulsar SAX\,J1808.4$-$3658.\ \textit{\mnras} \textbf{343}, 1301--1311 (2003).

\bibitem{Romani} Romani, R.~W.\ Gamma-Ray Pulsars: Radiation Processes in the Outer Magnetosphere.\ \textit{Astrophys. J.} \textbf{470}, 469 (1996).

\bibitem{Torres2019} Torres, D.~F., Vigan{\`o}, D., Coti Zelati, F. \& Li, J.\ Synchrocurvature modelling of the multifrequency non-thermal emission of pulsars.\  \textit{\mnras} \textbf{489}, 5494--5512 (2019).

\bibitem{Mignani2019} Mignani, R.~P. \textit{et al.}\ The First Ultraviolet Detection of the Large Magellanic Cloud Pulsar PSR\,B0540$-$69 and Its Multi-wavelength Properties.\ \textit{Astrophys. J.} \textbf{871}, 246 (2019).

\end{thebibliography}

\subsection{Data availability.}
The barycentered SiFAP2 data that support the findings of this study are available in figshare at \url{http://dx.doi.org/10.6084/m9.figshare.12707444}. 

\section*{Reference list.}


\begin{addendum}
\item A.M.Z. thanks the HST Contact Scientist, D. Welty (STScI), for constant support in the observation planning and T. Royle (STScI) for checking the scheduling processes. A.M.Z. thanks A. Riley (STIS Team) for the support in the scientific data analysis. A.M.Z. acknowledges the support of the PHAROS COST Action (CA16214) and A. Ridolfi for his help in data analysis. A.M.Z. would also like to thank G. Benevento for comments on draft. 
F.C.Z is supported by a Juan de la Cierva fellowship. S.C. and P.D.A. acknowledge support from ASI grant I/004/11/3.
D.d.M., A.P. and L.S. acknowledge financial support from the Italian Space Agency (ASI) and National Institute for Astrophysics (INAF) under agreements ASI-INAF I/037/12/0. L.B., D.d.M., T.D.S., A.P. and L.S. acknowledge financial contributions from  ASI-INAF agreement n.2017-14-H.0, INAF main-stream (PI: T. Belloni; PI: A. De Rosa) 
D.F.T. acknowledges support from grants PGC2018-095512-B-I00, SGR2017-1383, and AYA2017-92402-EXP.
L.B and T.d.S. thank A. Marino for useful discussions and acknowledge financial contributions from the HERMES project financed by the Italian Space Agency (ASI) Agreement n. 2016/13 U.O. 
T.d.S and L.S. acknowledge the iPeska research grant (PI: Andrea Possenti) funded under the INAF national call Prin-SKA/CTA approved with the Presidential Decree 70/2016. A.P., F.C.Z., and D.T.
acknowledge the International Space Science Institute (ISSI-Beijing), which funded and
hosted the international team ‘Understanding and Unifying the Gamma-rays Emitting
Scenarios in High Mass and Low Mass X-ray Binaries’.
Results obtained with SiFAP2 and presented in this paper are based on observations made with the Italian {\it Telescopio Nazionale Galileo} (TNG) operated by the {\it Fundaci\'on Galileo Galilei} (FGG) of the
{\it Istituto Nazionale di Astrofisica} (INAF) at the {\it  Observatorio del Roque de los Muchachos} (La Palma, Canary Islands, Spain). 
Part of this paper is based on observations with the NASA/ESA Hubble Space Telescope, obtained at the Space Telescope Science Institute, which is operated by AURA, Inc., under NASA contract NAS5-26555.
This work also made use of data and software provided by the High Energy Astrophysics Science Archive Research Center (HEASARC).
 
\item [Author contributions] F.A. and A.M.Z. contributed with equal proportion to this study. F.A., A.M.Z., A.P. and F.C.Z. analysed optical, ultraviolet and X-ray data. F.A., A.M.Z., A.P., F.C.Z, and L.S. wrote the paper. A.M.Z., A.P., S.C., P.D.A., F.C.Z., P.C., L.S, T.D.S, L.B, D.d.M., D.F.T., G.L.I., and A.S. interpreted the results. F.A., F.M., P.Cr., A.G., F.L., and E.P. conceived the SiFAP2. A.G., A.P., F.A. performed the optical observation. A.G., M.Ce., M.D.G.G., A.L.R.R., H.P.V., M.H.D., and J.J.S.J. developed the SiFAP2 mechanical interface and its relative control software. M.C. and R.P.M contributed to the HST data analysis. M.C.B., D.M.R, D.M.B, and F.L. contributed to the optical part of the SED. All authors read, commented on and approved submission of this article.  

\item [Competing Interests] The authors declare that they have no
competing financial interests.
\item [Correspondence] Correspondence and requests for materials
should be addressed to F.A. and A.M.Z. (email: filippo.ambrosino@inaf.it, arianna.miraval@inaf.it).
\end{addendum}


\end{document}